\documentclass[12pt,preprint]{aastex}

\shorttitle{Rotation of Janus and Epimetheus}
\shortauthors{Noyelles}

\usepackage{lscape}
\usepackage{wasysym}
\begin{document}


\title{Theory of the rotation of Janus and Epimetheus}


\author{Beno\^it Noyelles}
\affil{University of Namur, Dpt of Mathematics, 8 Rempart de la Vierge,
B-5000 Namur, Belgium \and
IMCCE, CNRS UMR 8028, Paris Observatory, UPMC, USTL, 77 avenue Denfert-Rochereau, 75014 Paris, France}
\email{noyelles@imcce.fr}

%
%




\begin{abstract}

\par The Saturnian coorbital satellites Janus and Epimetheus present a unique dynamical configuration in the Solar System, because of high-amplitude horseshoe orbits, due to a mass ratio of order unity. As a consequence, they swap their orbits every 4 years, while their orbital periods is about $0.695$ days. Recently, \citet{ttb2009} got observational informations on the shapes and the rotational states of these satellites. In particular, they detected an offset in the expected equilibrium position of Janus, and a large libration of Epimetheus.

\par We here propose to give a 3-dimensional theory of the rotation of these satellites in using these observed data, and to compare it to the observed rotations. We consider the two satellites as triaxial rigid bodies, and we perform numerical integrations of the system in assuming the free librations as damped.

\par The periods of the three free librations we get, associated with the 3 dimensions, are respectively $1.267$, $2.179$ and $2.098$ days for Janus, and $0.747$, $1.804$ and $5.542$ days for Epimetheus. The proximity of $0.747$ days to the orbital period causes a high sensitivity of the librations of Epimetheus to the moments of inertia. Our theory explains the amplitude of the librations of Janus and the error bars of the librations of Epimetheus, but not an observed offset in the orientation of Janus.

\end{abstract}





\section{Introduction}

\par The Saturnian system has many coorbital satellites, like Tethys-Telesto-Calypso and Dione-Helen-Polydeuces, but the pair Janus-Epimetheus 
presents a unique configuration because their not-so-big mass ratio, i.e. $\approx3.604$ \citep{jspbcem2008}, induces large librations named
horseshoe orbits \citep{dm1981}. These inner satellites have originally been observed by Dollfus in 1966 \citep{d1967}, who thought to have
 discovered only one satellite, named Janus, but found it difficult, if not impossible, to fit a reliable orbit of this body 
to observations. Ten years later, \citet{fl1977} showed that this discrepancy could be explained by the presence of at least one another satellite, 
now known as Epimetheus, whose existence was confirmed in 1980 thanks to Earth-based observations and Voyager images (see e.g. \citet{lsfr1981}).

\par The horseshoe orbits are the consequence of a $1:1$ orbital resonance (i.e., Janus and Epimetheus have the same mean orbital period), that provokes an orbital swap every 4 years, the two satellite recovering their original semimajor axis after a second swap. From a dynamical point a view, we can say that the period of libration associated with this $1:1$ resonance is 8 years.

\par The Cassini spacecraft, currently orbiting in the Saturnian system, presents a unique opportunity to observe these orbital swaps and 
their consequences, for instance on the density waves in the rings (see e.g. \citet{tnbhp2006}). Recently, Cassini images of Janus and Epimetheus allowed
to derive the shapes of these bodies, an estimation of their moments of inertia, and also some measures of their rotation. In particular, it has been 
established that there is a permanent offset of several degrees between Janus minimum moment of inertia and the equilibrium sub-Saturn point, and that 
Epimetheus has an oscillation about synchronous rotation of $5.9^{\circ}\pm1.2^{\circ}$ \citep{ttb2009}. These last two results are expected to give 
clues on the distribution of masses inside of these two bodies, as did for instance the detection of an oscillation of Mercury around the $3:2$ orbital
resonance \citep{mpjsh2007}, considered as the evidence of a molten core.

\par The goal of this paper is to compute a theory of the rotation of Janus and Epimetheus, and to compare it with the observational data. We first review some aspects of the orbital dynamics of Janus and Epimetheus, because they have a direct influence on the rotation through the Saturnian perturbation. In particular, we express the fundamental frequencies of the orbital perturbations, and interpolate JPL/HORIZONS ephemerides before including them into a 3-degrees of freedom numerical model of the rotation of Janus and Epimetheus, seen as rigid bodies. Finally, we propose some interpretations of the observed data, in comparison with our theory.

\section{An analysis of the orbital motion}

\par It is necessary to know as accurately as possible the orbital motion of Janus and Epimetheus about Saturn, to estimate the influence of the Saturnian torque on their rotation. The most reliable ephemerides of Janus and Epimetheus available are JPL HORIZONS' ones \footnote{http://ssd.jpl.nasa.gov/?horizons}, elaborated by \citet{jspbcem2008}. This server provides data tables of the positions, velocities, orbital elements,\ldots of the most important Solar System objects (including Janus and Epimetheus) at given dates. We explain in this part how we extract from JPL data the useful informations that will help us to understand the orbital dynamics of these bodies.

\subsection{The proper modes}

\par The first information we are interested in is the proper modes of the orbital motion, i.e. the fundamental frequencies of the system. If we assume that the orbital motion of a given body is not chaotic, its variables can be expressed under a synthetic form, i.e. an infinite sum of sinusoidal terms, that depend on a limited number of fundamental frequencies, this number depending itself on the number of involved bodies. If we neglect the periodic influence of the other Saturnian satellites and consider that our system is only composed of an oblate Saturn, Janus and Epimetheus, we can restrict our fundamental frequencies to 6 elements, i.e. 3 for each satellites, these frequencies being the orbital one, and the slow frequencies of precessions of nodes and pericentres. The 1:1 orbital resonance forces the equality of the two mean motions, but also results in the apparition of a proper frequency associated with this resonance, of period close to 8 years, i.e. two orbital swaps. So we have 6 proper modes.

\par From the JPL cartesian coordinates (NAIF kernels SAT299 for Janus and Epimetheus, and SAT317 for Saturn) and masses of Janus and Epimetheus, we derived the following keplerian elements, in the reference frame centered on the center of mass of Saturn, and referring to the equatorial plane of Saturn and the node of this plane with the ecliptic at J2000.0:

\begin{itemize}

\item $a$: semi-major axis

\item $\lambda$: mean longitude

\item $z=k+\sqrt{-1}h=e\exp\big(\sqrt{-1}\varpi\big)$, $e$ and $\varpi$ being respectively the eccentricity and the longitude of the pericentre

\item $\zeta=q+\sqrt{-1}p=\sin\Big(\frac{I}{2}\Big)\exp\big(\sqrt{-1}\ascnode\big)$, $I$ and $\ascnode$ being respectively the inclination and the 
longitude of the ascending node.
 
\end{itemize}

\par These elements are the classical elliptic ones. Another possibility could be to use the epicyclic elements. The basic idea is to consider the oblateness of Saturn in the computation of the osculating elements. The third Kepler law is being modified this way:

\begin{equation}
n^2a^3=GM_{\saturn}\Bigg(1+\frac{3}{2}J_2\Bigg(\frac{R_{\saturn}}{a}\Bigg)^2+O(J_4)\Bigg),
\end{equation}
$n$ being the instantaneous mean motion of the considered body (Janus or Epimetheus), $M_{\saturn}$ Saturn's mass, and $R_{\saturn}$ its equatorial radius (see e.g. \citet{g1981}). This formulation does not change the fundamental frequencies, because they physically correspond to revolutions of the body about its parent planet, but change the semimajor axes, eccentricities and inclinations. In particular, it can be shown that it drastically reduces the amplitudes of the short period oscillations in the case of a small body orbiting close to an oblate planet \citep{b1959}. Such a formulation is very convenient for describing the dynamics of planetary rings \citep{bl1987,lb1991,bl1994,rs2006} because it gives quite constant elements and a smaller eccentricity than the classical elliptic elements, but in our case we want to identify the mean orbital period. Moreover, no influence is expected on the representation of the orbital swap. This is the reason why we did not use them.

\par Once these elements have been obtained as data tables, we performed frequency analyses to get periodic time series.  The frequency analysis algorithm 
we used is based on Laskar's original idea, named NAFF as Numerical Analysis of the Fundamental Frequencies (see for instance \citet{l1993} for the method, and \citet{l2003} for the convergence proofs). It aims at identifying the coefficients $a_k$ and $\omega_k$ of a complex signal $f(t)$ obtained numerically over a finite time span $[-T;T]$  and verifying

\begin{equation}
\label{equ:naff}
f(t) \approx \sum_{k=1}^na_k\exp(\sqrt{-1}\omega_kt),
\end{equation}
where $\omega_k$ are real frequencies and $a_k$ complex coefficients. If the signal $f(t)$ is real, its frequency spectrum is symmetric and the complex amplitudes associated with the frequencies $\omega_k$ and $-\omega_k$ are complex conjugates. The frequencies and amplitudes associated are found with an iterative scheme. To determine the first frequency $\omega_1$, one searches for the maximum of the amplitude of 

\begin{equation}
\label{equ:philas}
\phi(\omega)=<f(t),\exp(\sqrt{-1}\omega t)>,
\end{equation}
where the scalar product $<f(t),g(t)>$ is defined by

\begin{equation}
\label{equ:prodscal}
<f(t),g(t)>=\frac{1}{2T}\int_{-T}^T f(t)\overline{g(t)}\chi(t) dt,
\end{equation}
and where $\chi(t)$ is a weight function, i.e. a positive function with

\begin{equation}
\label{equ:poids}
\frac{1}{2T}\int_{-T}^T \chi(t) dt=1.
\end{equation}
Once the first periodic term $\exp(\sqrt{-1}\omega_1t)$ is found, its complex amplitude $a_1$ is obtained by orthogonal projection, and the process is started again on the remainder $f_1(t)=f(t)-a_1\exp(\sqrt{-1}\omega_1t)$. The algorithm stops when two detected frequencies are too close to each other, what alters their determinations, or when the number of detected terms reaches a maximum set by the user. This algorithm is very efficient, except when two frequencies are too close to each other. In that case, the algorithm is not confident in its accuracy and stops. When the difference between two frequencies is larger than twice the frequency associated with the length of the total time interval, the determination of each fundamental frequency is not perturbed by the other ones. Although the iterative method suggested by \citet{c1998} allows to reduce this distance, some troubles still remain when the frequencies are too close to each other.

\par We used ephemerides over the time interval 1950-2050, i.e. the widest interval over which the JPL ephemerides are available. This time span is quite long in comparison with the orbital period of these two satellites ($\approx 17$ hours), and long enough for estimating the period of the resonant argument ($\approx 8$ years, i.e. two orbital swaps). The shortest unaliased period that can be detected is twice the time step. Here our time step was 3 hours, what is short enough to detect 8h-contributions, i.e. half the orbital period expected.

\par The frequency analyses of the keplerian elements of Janus and Epimetheus allowed us to identify 7 proper modes (see Tab.\ref{tab:propmod}), 
written as $\lambda$, $\phi$, $\varpi_J$, $\varpi_E$, $\ascnode_J$, $\ascnode_E$ and $\omega$. The first six proper modes are the ones expected, while the last one, $\omega$, has been actually detected in the frequency analyses. This seventh proper mode is difficult to identify, particularly because its period is larger than our time interval, so we lack of accuracy on its period. Moreover, the amplitude associated is small. $\lambda$ is close to the mean longitude of Janus, $\phi$ is the resonant argument, $\varpi_J$ and $\varpi_E$ are close to the longitudes of the pericentres of Janus and Epimetheus, while $\ascnode_J$ and $\ascnode_E$ are close to the longitudes of their ascending nodes. By "close to", we mean that the proper mode is the main component of the keplerian element associated with.


\begin{table}[ht]
 \centering
\caption[Proper modes of the orbital motions]{Proper modes of the orbital motions of Janus and Epimetheus. The column "Origin" gives the variable from which the given numerical values of the proper mode have been extracted, while the last column gives the frequencies given by \citet{jspbcem2008}, fitted on $[2003;2005]$. \label{tab:propmod}}
\begin{tabular}{c|rrrr|r}
\tableline\tableline
 & Frequency ($rad/y$) & Phase at J2000 & Period & Origin & JSP2008 \\
\tableline
$\lambda$ &      $3304.0143278$ & $-114.564^{\circ}$ &     $0.69459$ d & $\lambda_J$ & $3303.6716$ \\
 & & & & $\lambda_E$ & $3305.2554$ \\
$\phi$ &            $0.7847244$ &  $177.674^{\circ}$ &     $8.00687$ y & $a_J$ & -- \\
$\varpi_J$ &       $13.0908741$ &  $129.064^{\circ}$ &   $175.30788$ d & $z_J$ & $13.0869$ \\
$\varpi_E$ &       $13.0928523$ & $-121.751^{\circ}$ &   $175.28140$ d & $z_E$ & $13.1022$ \\
$\ascnode_J$ &    $-13.0386776$ &  $114.152^{\circ}$ &  $-176.00968$ d & $\zeta_J$ & $-13.0359$\\
$\ascnode_E$ &    $-13.0400438$ &  $152.811^{\circ}$ &  $-175.99124$ d & $\zeta_E$ & $-13.0512$ \\
$\omega$ &          $0.0461439$ & $-120.692^{\circ}$ &   $136.16498$ y & $\zeta_J$ & --
\end{tabular}
\end{table}

\par The numerical values we give in this table come from outputs of the frequency analyses. We can see that some frequencies are very close to each other, for instance for the precessions of the pericentres. They could be distinguished from each others in the same signal if the interval of study were longer than twice the period associated with the difference of their frequencies, i.e. $\approx$ 200,000 years. Nevertheless, we can also use the phases to identify the contribution of each proper mode, that is the reason why we are confident in their identification. Our numerical frequencies are in good agreement with the ones given by \citet{jspbcem2008}, that have been obtained from a fit over two years.

\par We here give the frequency analyses of the two semi-major axes (Tab.\ref{tab:smajan} for Janus and Tab.\ref{tab:smaepim} for Epimetheus). The decomposition of the other orbital variables can be found in the supplemental material. These tables should be used this way: the Tab.\ref{tab:smajan} means that the semimajor axis of Janus could be approximated by

\begin{eqnarray}
a(t) & \approx & 152043.049 \nonumber \\
 & + & 13.182\cos(\phi(t)) \nonumber \\
 & + & 7.397\cos(\lambda(t)-\varpi_J(t)) \nonumber \\
 & - & 4.538\cos(3\phi(t)) \nonumber \\
 & + & \ldots
\end{eqnarray}
with $\phi(t)=0.7847244t+177.674^{\circ}$, $\lambda(t)=3304.0143278t-114.564^{\circ}$ and $\varpi_J(t)=13.0908741t+129.064^{\circ}$ (Tab.\ref{tab:propmod}), the amplitudes being in km, the frequencies in rad/y, and the time in years, the origin being J2000. The cosines should be replaced by sines or by complex exponentials when it is stated in the caption. The cosines and sines are suitable for real variables, and the opportunity to use sine or cosine is decided in reading the phases.

\begin{table}[ht]
 \centering
\caption[Semi-major axis of Janus]{Semi-major axis of Janus. The series are in cosine.\label{tab:smajan}}
\begin{tabular}{r|rrrrrr}
\tableline\tableline
 N & $\lambda$ & $\phi$ & $\varpi_J$ & Amplitude (km) & Period \\
\tableline
 $1$ & -   & -     & -    & $152043.049$ & $\infty$ \\
 $2$ & -   & $1$   & -    &     $13.182$ &   $8.00687$ y \\
 $3$ & $1$ & -     & $-1$ &      $7.397$ &   $0.69735$ d \\
 $4$ & -   & $3$   & -    &     $-4.538$ &   $2.66892$ y \\
 $5$ & -   & $5$   & -    &      $2.403$ &   $1.60135$ y \\
 $6$ & $1$ & $-1$  & $-1$ &      $2.098$ &   $0.69752$ d \\
 $7$ & $1$ & $1$   & $-1$ &      $2.097$ &   $0.69719$ d \\
 $8$ & -   & $7$   & -    &     $-1.547$ &   $1.14382$ y \\
 $9$ & -   & $9$   & -    &      $1.087$ & $324.94105$ d \\
$10$ & -   & $11$  & -    &     $-0.802$ & $265.86068$ d \\
$11$ & -   & $13$  & -    &      $0.611$ & $224.95882$ d \\
$12$ & -   & $15$  & -    &     $-0.476$ & $194.96413$ d \\
$13$ & -   & $17$  & -    &      $0.377$ & $172.02726$ d \\
$14$ & $1$ &  $2$  & $-1$ &      $0.365$ &   $0.69702$ d \\
$15$ & $1$ & $-2$  & $-1$ &      $0.359$ &   $0.69769$ d \\
$16$ & -   & $19$  & -    &     $-0.302$ & $153.91905$ d \\
$17$ & -   & $21$  & -    &      $0.245$ & $139.26011$ d \\
$18$ & $1$ & $-3$  & $-1$ &      $0.203$ &   $0.69785$ d \\
$19$ & -   & $23$  & -    &     $-0.201$ & $127.15028$ d \\
$20$ & $1$ &  $3$  & $-1$ &      $0.200$ &   $0.69685$ d \\
$21$ & -   & $25$  & -    &      $0.165$ & $116.97818$ d \\
\tableline
\end{tabular}
\end{table}

\begin{table}[ht]
 \centering
\caption[Semi-major axis of Epimetheus]{Semi-major axis of Epimetheus. The series are in cosine.\label{tab:smaepim}}
\begin{tabular}{r|rrrrr}
\tableline\tableline
 N & $\lambda$ & $\phi$ & $\varpi_E$ & Amplitude (km) & Period \\
\tableline
 $1$ & -   & -     & -    & $152043.602$ & $\infty$ \\
 $2$ & -   &  $1$  & -    &    $-47.500$ &   $8.00675$ y \\
 $3$ & -   &  $3$  & -    &     $16.353$ &   $2.66892$ y \\
 $4$ & -   &  $5$  & -    &     $-8.660$ &   $1.60135$ y \\
 $5$ & $1$ & $-1$  & $-1$ &     $-6.251$ &   $0.69752$ d \\
 $6$ & $1$ &  $1$  & $-1$ &     $-6.234$ &   $0.69719$ d \\
 $7$ & -   &  $7$  & -    &      $5.577$ &   $1.14382$ y \\
 $8$ & $1$ &  $2$  & $-1$ &     $-4.731$ &   $0.69702$ d \\
 $9$ & $1$ & $-2$  & $-1$ &     $-4.726$ &   $0.69735$ d \\
$10$ & -   &  $9$  & -    &     $-3.918$ & $324.94138$ d \\
$11$ & $1$ & -     & $-1$ &     $-3.114$ &   $0.69735$ d \\
$12$ & -   & $11$  & -    &      $2.889$ & $265.86079$ d \\
$13$ & -   & $13$  & -    &     $-2.201$ & $224.95893$ d \\
$14$ & -   & $15$  & -    &      $1.715$ & $194.96444$ d \\
$15$ & -   & $17$  & -    &     $-1.359$ & $172.02730$ d \\
$15$ & $1$ &  $3$  & $-1$ &     $-1.319$ &   $0.69685$ d \\
$16$ & $1$ & $-3$  & $-1$ &     $-1.293$ &   $0.69785$ d \\
$17$ & -   & $19$  & -    &      $1.090$ & $153.91905$ d \\
\tableline
\end{tabular}
\end{table}

\par A striking point in these decompositions are the odd harmonics of $\phi$. Their presence can be explained in considering that the semimajor axes are close to square wave, and that the Fourier series of a square wave is composed only of odd harmonics. We can also notice the quite slow decrease of the amplitude, showing that the quasi-periodic decomposition of such a signal converges slowly. Here, all the terms detected by the frequency analysis are given, there is no cut-off based on the amplitude. The estimation of the accuracy of this representation will be given later on the positions of the bodies (Fig.\ref{fig:distance}).

\par In addition to these frequency analyses, we also find drifts in the orbital elements of Janus and Epimetheus, indicating the presence of long-term effects, that a representation over one century cannot render. We made a linear least-squares fit of the eccentricities $e$ and inclinations $I$ of Janus and Epimetheus and we get, for Janus, $e=(3.93785\times10^{-6}\pm1.708\times10^{-7})t+7.30539\times10^{-3}\pm4.93\times10^{-6}$, and $I=((-9.31642\times10^{-4}\pm5.263\times10^{-6})t+9.86587\pm1.519\times10^{-4})$ arcmin, and for Epimetheus $e=(-1.03189\times10^{-5}\pm1.74\times10^{-7})t+1.01621\times10^{-2}\pm5.023\times10^{-6}$ and $I=((1.49189\times10^{-3}\pm5.529\times10^{-6})t+21.1834\pm1.596\times10^{-4})$ arcmin, the time unit being the year and its origin J2000.0.

\subsection{Focus on the orbital swap}

\par The orbital swap occurring every 4 years is the key point of the dynamics of Janus and Epimetheus. We here propose to recall its main aspects, that have been extensively studied in previous works, e.g. \citet{ycsy1983,yss1989,nhmy1992}.


\begin{figure}[ht]
\centering
\begin{tabular}{cc}
 \includegraphics[height=5cm,width=7.8cm]{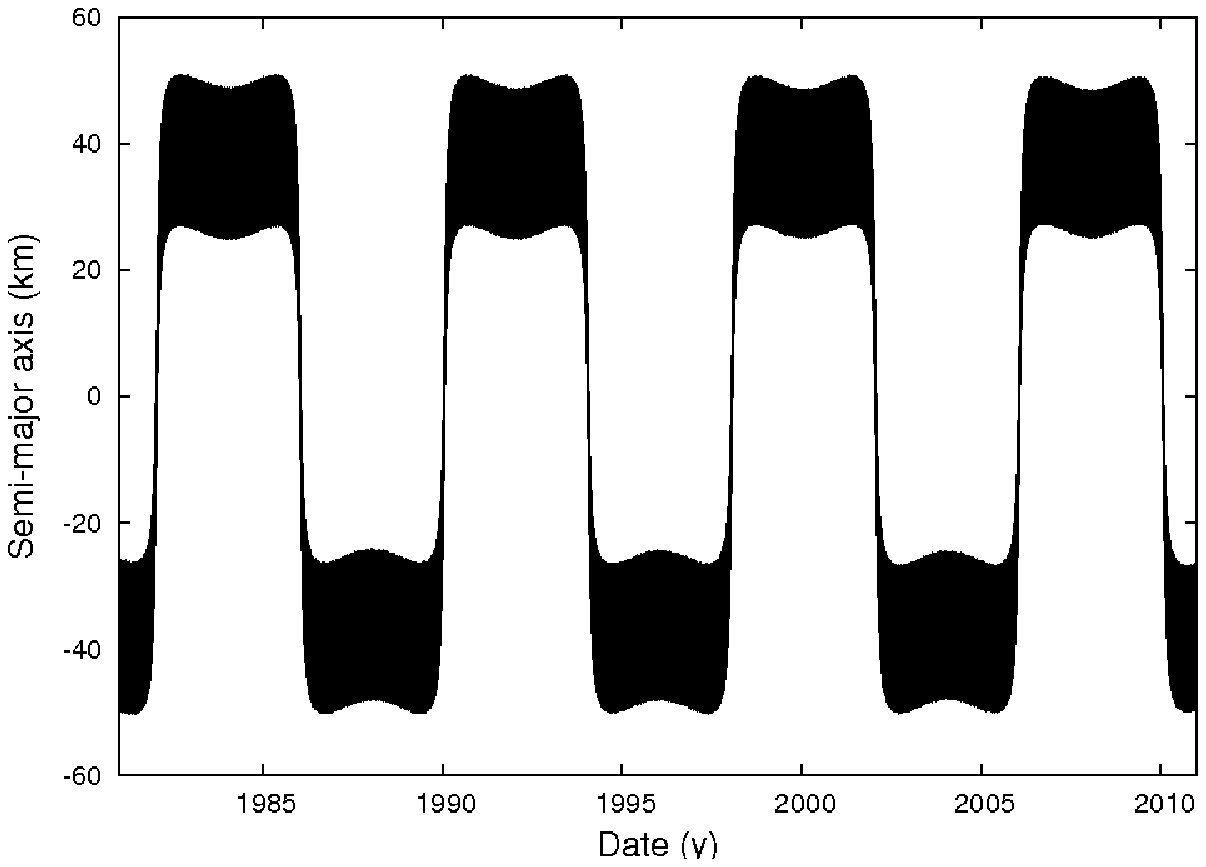} & \includegraphics[height=5cm,width=7.8cm]{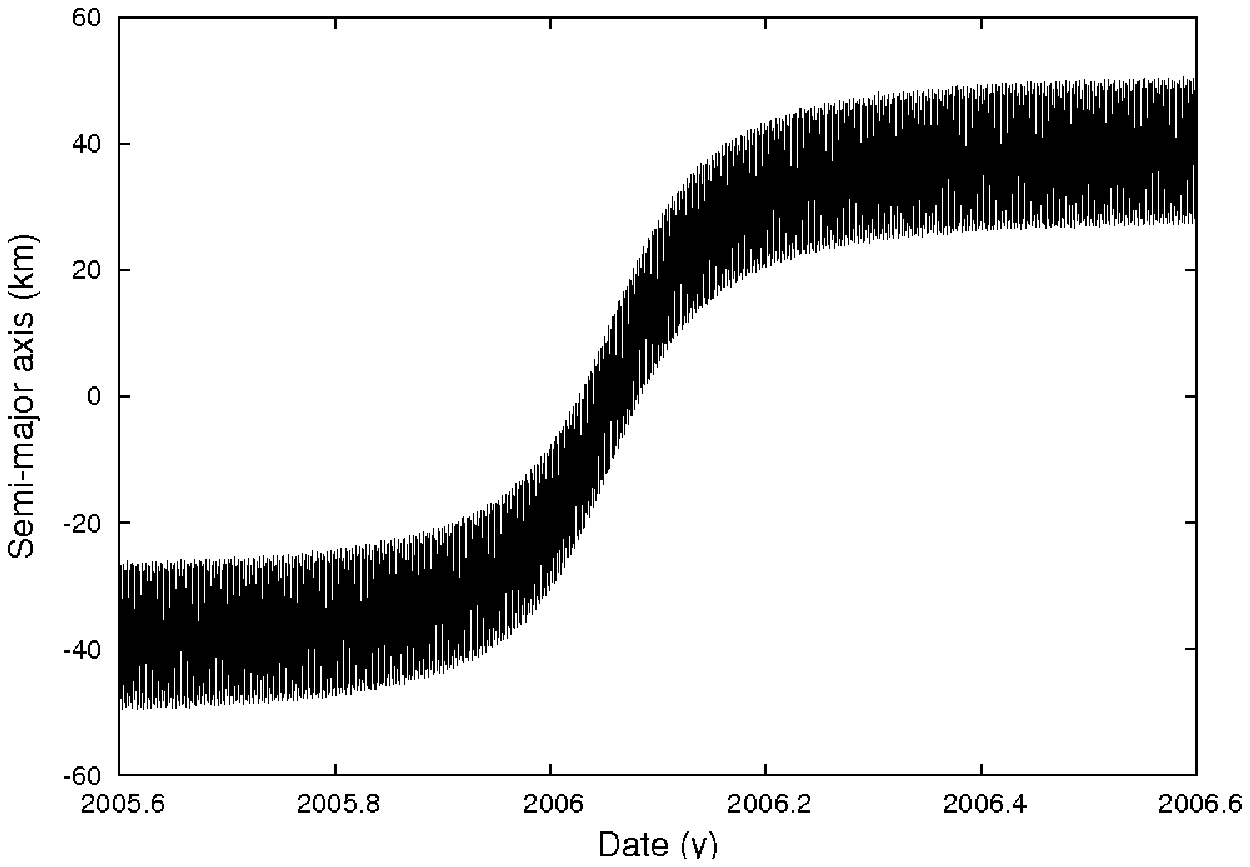} \\
(a) & (b) \\
 \includegraphics[height=5cm,width=7.8cm]{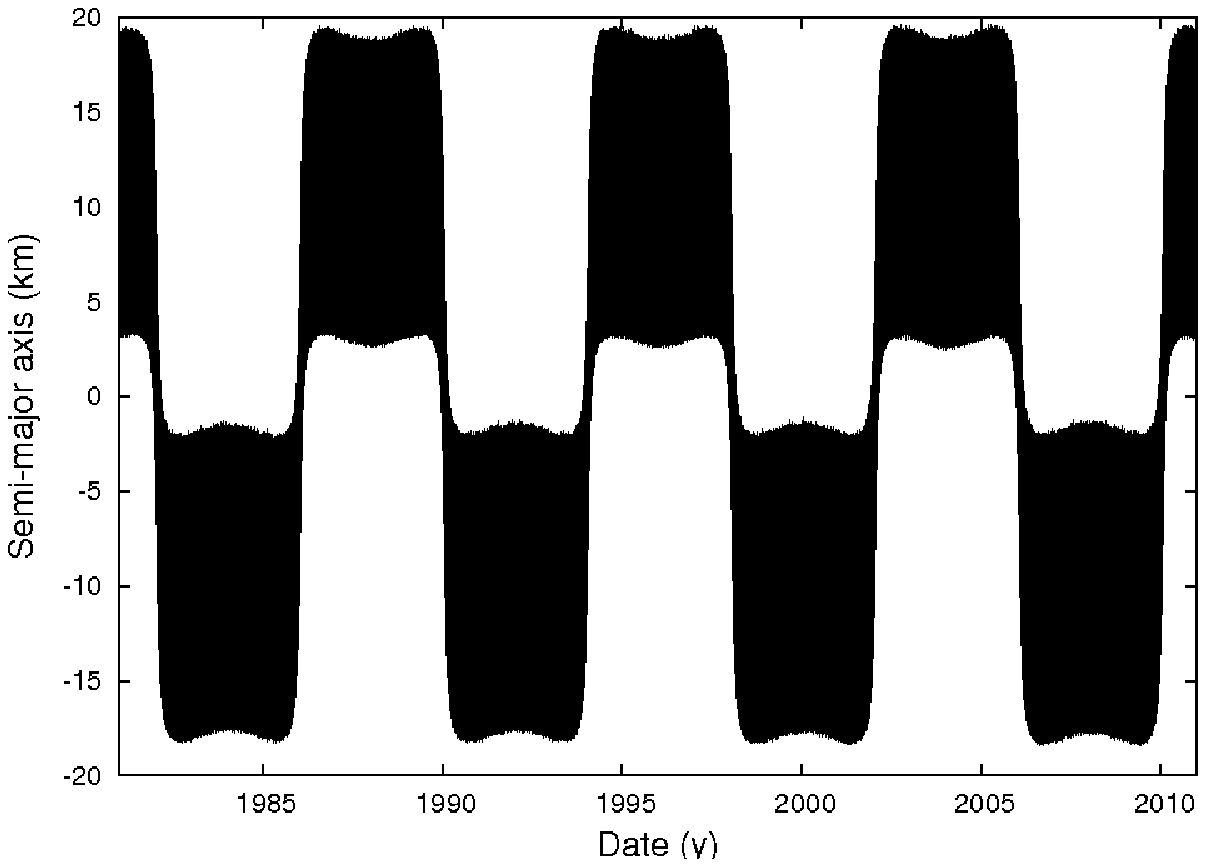} & \includegraphics[height=5cm,width=7.8cm]{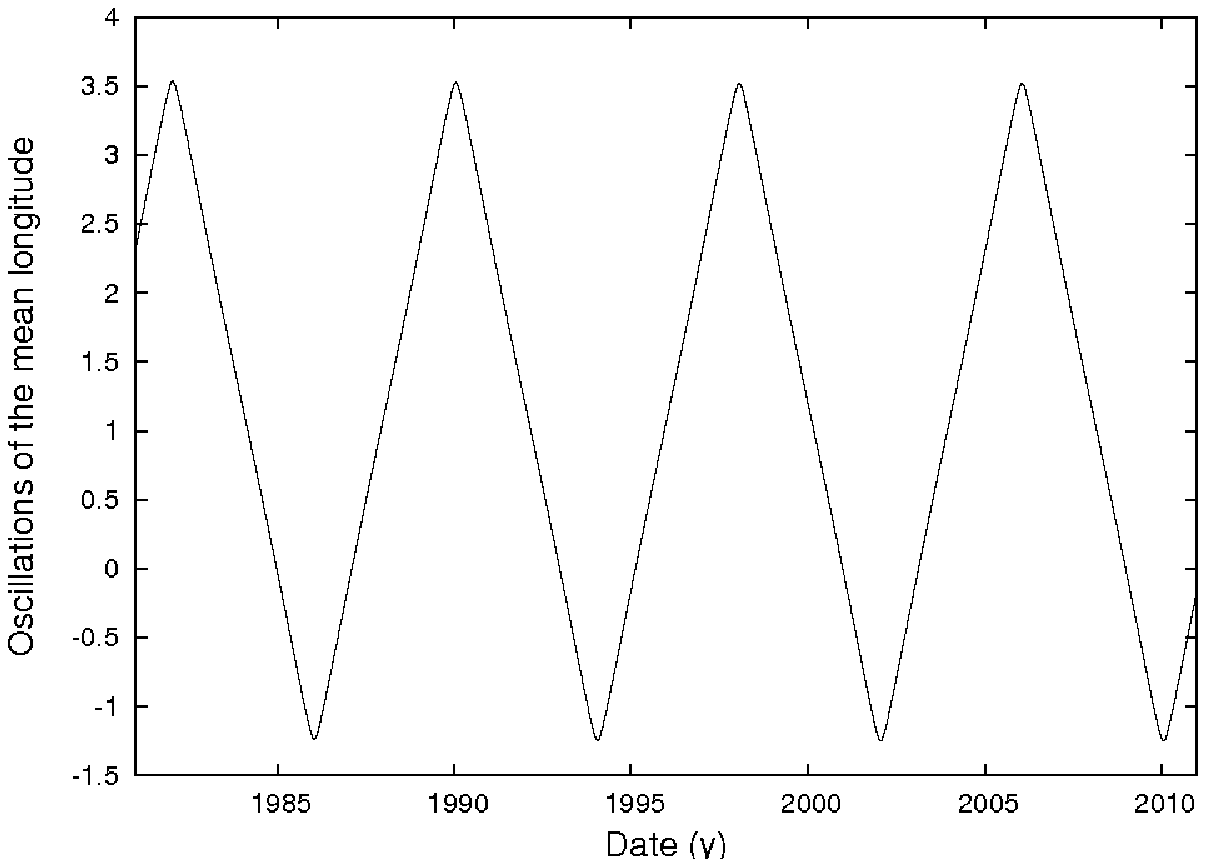} \\
(c) & (d)
\end{tabular}
\caption[The orbital swap]{The orbital swap, plotted with JPL HORIZONS data. The panels (a) and (b) show the variations of the semi-major axis of Epimetheus, while the panel (c) shows the one of Janus. The panel (d) represents the variations of the mean longitude of Epimetheus after removal of a slope of $3304.01449932$ rad/y. \label{fig:swap}}
\end{figure}

\par The Figure \ref{fig:swap} shows the variations of the semi-major axes and the mean longitudes of Epimetheus because of the orbital swap. These 
figures have been obtained with the keplerian elements derived from the JPL HORIZONS cartesian coordinates. The most striking is the exact synchronization between the orbital swaps of the two bodies (panels (a) and (c)), illustrating the energy exchanges due to the $1:1$ orbital resonance. We can also notice the amplitudes of the swaps, respectively $\approx 13$ km for Janus and $\approx 47$ km for Epimetheus. The ratio between these two amplitudes is the mass ratio of the two bodies (i.e. $\approx 3.6$), as already shown by previous authors (e.g. \citet{dm1981}, \citet{ycsy1983}). We also notice the thickness of the plateau, due to short period oscillations whose amplitude partly depends on the orbital eccentricity of the considered body. As already said, this plateau would have been thinner with epicyclic elements. The consequences of the resonance on the mean longitudes are 8-y periodic sawtooth waves.

\par Finally,  the panel (b) is a zoom on the transition, similar to the Figure 2b of \citet{nhmy1992}. It illustrates the motion of Epimetheus during the swap. We can graphically evaluate the duration of the swap at about 6 months.

\subsection{Using the ephemerides in the numerical computations}

\par
\begin{figure}
 \epsscale{1.1}
 \plottwo{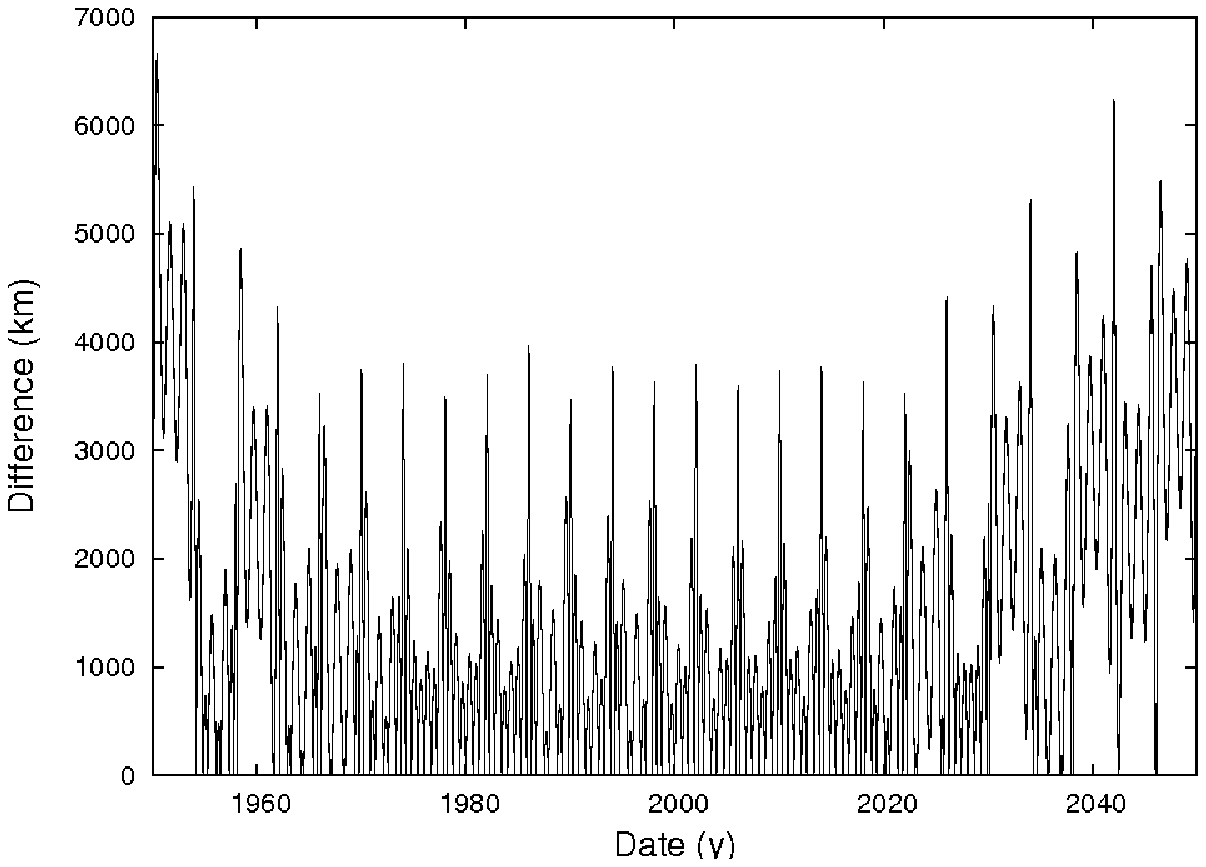}{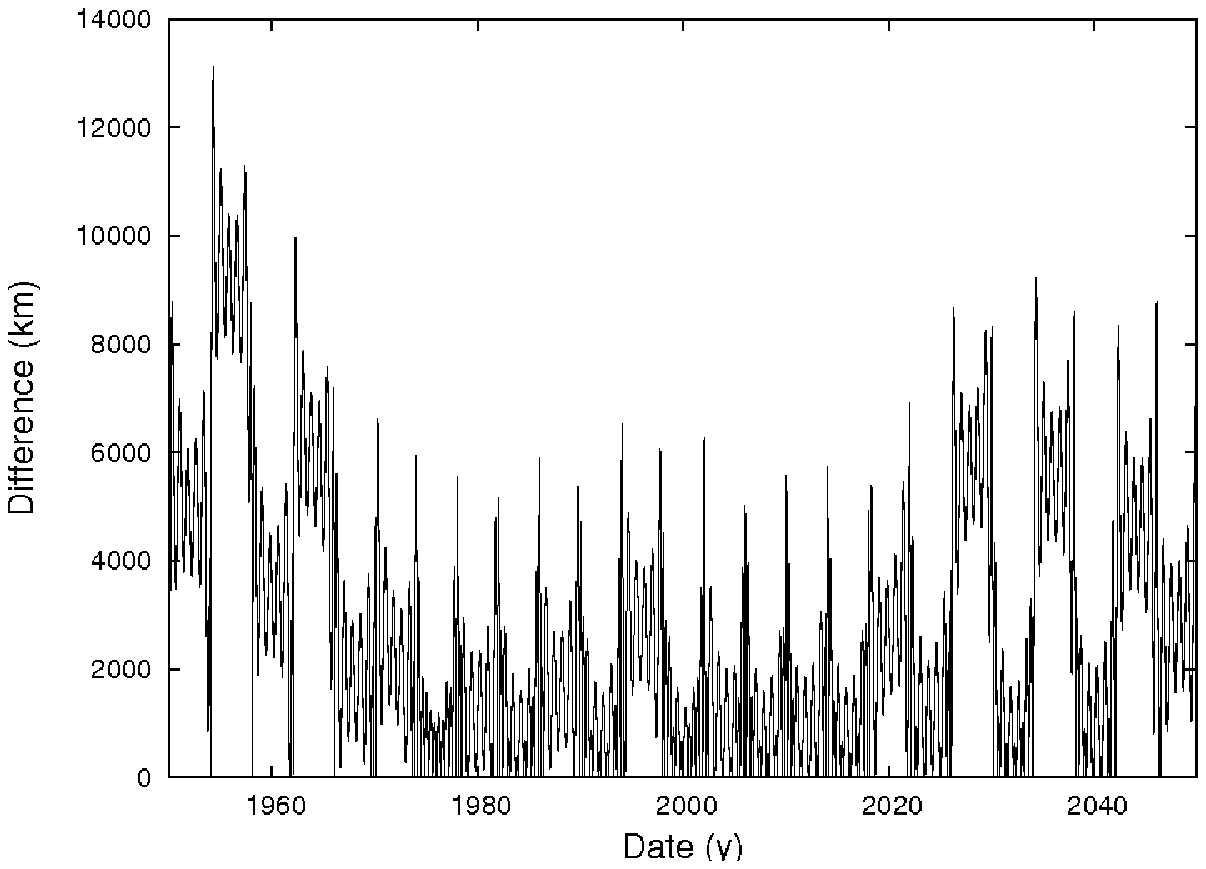}
 \caption{Distance between the positions given by JPL HORIZONS and those computed from our periodic time series of the orbital elements, for Janus (left) and Epimetheus (right). The error is due to the difficulty to represent these orbital elements with sinusoidal terms.\label{fig:distance}}
\end{figure}

 The quasi-periodic decomposition of the orbital ephemerides is a good way to get a description of the ephemerides that is physically understandable, but it can be reliably used in numerical computation only if its residuals with the real ephemerides are small enough. The Fig.\ref{fig:distance} shows them as distances between the positions of the satellites given by the JPL/HORIZONS ephemerides, and the quasiperiodic decompositions of the osculating elements, over the 100 years covered by HORIZONS ephemerides. We can see important residuals (up to 13,000 km for Epimetheus, while the least Janus-Epimetheus distance is about 10,000 km), with a 4-y periodic amplitude, due to the unusual shapes of the orbital elements because of the orbital swaps. They induce Gibbs phenomena, i.e. big residuals at the swaps. Moreover, there are still residuals far to the swaps, probably due to a lack of accuracy in the determination of the short-periodic contributions. These last oscillations should be smaller in epicyclic elements. So, we considered this quasiperiodic decomposition as not accurate enough to be used in numerical computations. We still are confident in the proper modes in Tab.\ref{tab:propmod}, this error just means that a large amount of periodic terms is required to get accurate enough ephemerides. Moreover, we cannot exclude the presence of long-period contributions, for which a polynomial interpolation would be more appropriate over this time interval.


\par We finally chose to interpolate the cartesian coordinates (i.e. positions and velocities) of the two satellites using the cubic splines interpolation. The reader can find explanations of this method in \citep{db2008}. The error associated is of fourth order \citep{s1969}, it means that a reduction of the sampling step of the signal by a factor $\alpha$ will reduce the interpolation error by a factor $\approx \alpha^4$. For this, we used the interpolation routines given by the GNU Scientific Library \citep{gdtgjabr2009}, and checked the error in comparing the positions given by the interpolation routine with positions given by HORIZONS, at dates that were not used to interpolate the ephemerides (at the interpolation points, the error should be null) and got an interpolation maximum error of about $850$ km with a sampling step of 3 hours. We finally used a sampling step of 1 hour to get a maximum error of 9 km. Contrary to the quasiperiodic decomposition, the error seems to have a quite uniform repartition, in particular the dates of the orbital swaps cannot be guessed from the plots of errors. We do not show it here because the reader would just see black rectangles.

\par This accuracy of 9 km assumes that the JPL ephemerides are exact over [1950-2050]. Their internal error can have at least two causes. The first one is the interpolation error of the numerical integration used to compute the ephemerides. This error is considered as lower than 25 meters for the DE406 planetary ephemerides, and should be lower for the satellites of Saturn, so it is negligible in our study. The other cause of error comes from the fit of the dynamical model to the astrometric observations.  As explained in \cite{jspbcem2008}, this fit has been made mostly on Cassini observations since 2004, with a $1-\sigma$ accuracy of about 20 km in the downtrack direction. Out of this timespan (i.e. before 2004 and in the future), the accuracy is expected to be worse because of the extrapolation. The reader can find in \citet{daalv2009} a review of the methods used to estimate the accuracy of extrapolated ephemerides.

\section{The rotation}

\par We here use our interpolation of the JPL HORIZONS ephemerides to get the gravitational perturbation of Saturn on the rotation of Janus and Epimetheus, with a 1-hour time step and so an internal accuracy of 9 km. The influence of the other satellites is accounted only as indirect effects, i.e. through perturbations on the orbits of Janus and Epimetheus. As for most of the natural satellites of the Solar System, these two bodies are expected to follow the 3 Cassini Laws, originally enounced for the Moon \citep{c1693,c1966}, i.e.:

\begin{enumerate}

\item The Moon rotates uniformly about its polar axis with a rotational period equal to the mean sidereal period of its orbit about the Earth.

\item The inclination of the Moon's equator to the ecliptic is a constant angle (approximately $1.5^{\circ}$).

\item The ascending node of the lunar orbit on the ecliptic coincides with the descending node of the lunar equator on the ecliptic. This law could also be expressed as: the spin axis and the normals to the ecliptic and orbit plane remain coplanar.

\end{enumerate}
In the case of natural satellites, they can be rephrased this way: the rotation of the satellite is synchronous, its angular momentum has a nearly constant inclination on an inertial reference plane, and is located in the plane defined by the normal to the orbital plane and to the Laplace Plane. The Laplace Plane is the plane normal to the rotation axis of the orbital frame, i.e. it is defined with respect to the orbital precessional motion. It has the property to minimize the variations of the orbital inclinations. Using the Laplace Plane as the reference frame is of high importance for satellites orbiting far from their parent body, because of the Solar perturbation that tends to take the orbital plane away from the equatorial plane of the planet. However, for satellites orbiting close to their planet as it is the case here, the equatorial plane of Saturn is so close to the Laplace Plane that it can be used for describing the rotational dynamics (see \citet{d1993,n2009,ttn2009} and the Appendix).

\subsection{The model}

\par The model we used is very similar to the one already used by \citet{h2005io,h2005eu} for Io and Europa and by \citet{nlv2008} for Titan. We consider Janus and Epimetheus as rigid triaxial bodies whose matrices of inertia reads

\begin{equation}
I=\left(\begin{array}{ccc}
A & 0 & 0 \\
0 & B & 0 \\
0 & 0 & C
\end{array}\right)
\label{equ:inertie}
\end{equation}
with $A \leq B \leq C$.

\par The dynamical model is a 3-degree of freedom one in which 3 references frames are considered:

\begin{enumerate}

\item An inertial reference frame $(\vec{e_1},\vec{e_2},\vec{e_3})$. We used the one in which the orbital ephemerides are given, i.e. mean Saturnian equator and mean equinox for J2000.0 epoch.

\item A frame $(\vec{n_1},\vec{n_2},\vec{n_3})$ bound to the angular momentum of the considered body.

\item A frame $(\vec{f_1},\vec{f_2},\vec{f_3})$ rigidly linked to the body. 

\end{enumerate}


\begin{figure}
\epsscale{0.80}
\plotone{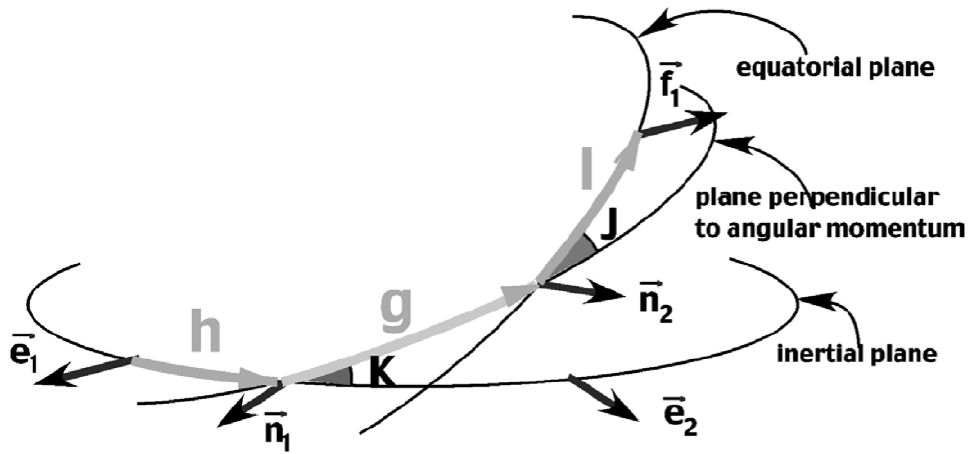}
\caption[The Andoyer variables]{The Andoyer variables (reproduced from \citep{h2005io}).\label{fig01}}
\end{figure}

\par We first use Andoyer's variables \citep{a1926,de1967}, which are based on two linked sets of Euler's angles. The first set $(h,K,g)$ locates the position of the angular momentum in the first frame $(\vec{e_1},\vec{e_2},\vec{e_3})$, while the second one, $(g,J,l)$, locates the body frame $(\vec{f_1},\vec{f_2},\vec{f_3})$ in the second frame tied to the angular momentum (see Fig. \ref{fig01}).

\par The canonical set of Andoyer's variables consists of the three angular variables $l,g,h$ and their conjugated momenta $L,G,H$ defined by the norm $G$ of the angular momentum and two of its projections : 

\begin{equation}
\begin{array}{lll}
l & \hspace{3cm} & L=G\cos J \\
g & \hspace{3cm} & G \\
h & \hspace{3cm} & H=G\cos K
\end{array}
\end{equation}

\par Unfortunately, these variables present two singularities: when $J=0$ (i.e., the angular momentum is colinear to $\vec{f_3}$), $l$ and $g$ are undefined, and when $K=0$ (i.e., when Janus/Epimetheus' principal axis of inertia is perpendicular to its orbital plane), $h$ and $g$ are undefined. That is the reason why we shall use the modified Andoyer's variables :

\begin{equation}
\begin{array}{lll}
p=l+g+h & \hspace{2cm} & P=\frac{G}{nC} \\
r=-h & \hspace{2cm} & R=\frac{G-H}{nC}=P(1-\cos K)=2P\sin^2\frac{K}{2} \\
\xi_q=\sqrt{\frac{2Q}{nC}}\sin q & \hspace{2cm} & \eta_q=\sqrt{\frac{2Q}{nC}}\cos q \label{equ:modified} \\
\end{array} \\
\end{equation}
where $n$ is the body's mean orbital motion , $q=-l$, and $Q=G-L=G(1-\cos J)=2G\sin^2\frac{J}{2}$. With these new variables, the singularity on $l$ has been dropped. Using these variables has a great mathematical interest, because they are canonical, so they simplify an analytical study of the system, as was done in previous works. Our study here is quite purely numerical, but we keep these variables, in order to be consistent with previous studies. We later derive other output variables, that are more relevant from a physical point of view.

\par In these variables, the kinetic energy $T=\frac{1}{2}\vec{\omega}\cdot\vec{G}$ of the system reads:

\begin{eqnarray}
\label{equ:kinrj}
T & = & \frac{nP^2}{2}+\frac{n}{8}\big[4P-\xi_q^2-\eta_q^2\big] \nonumber \\ 
& \times & \Big[\frac{\gamma_1+\gamma_2}{1-\gamma_1-\gamma_2}\xi_q^2+\frac{\gamma_1-\gamma_2}{1-\gamma_1+\gamma_2}\eta_q^2\Big]
\end{eqnarray}
with 

\begin{equation}
\gamma_1=\frac{2C-A-B}{2C}=J_2\frac{M\mathcal{R}^2}{C}
\label{equ:gama1}
\end{equation}
and

\begin{equation}
\gamma_2=\frac{B-A}{2C}=2C_{22}\frac{M\mathcal{R}^2}{C}.
\label{equ:gama2}
\end{equation}
In these last 3 formulae, $\vec{\omega}$ is the instantaneous vector of rotation, $M$ is the mass of Janus or Epimetheus, $\mathcal{R}$ its mean radius, and $J_2$ and $C_{22}$ the two classical normalized gravitational coefficients related respectively to the oblateness and equatorial ellipticity of the considered body.

\par The gravitational disturbing potential due to an oblate perturber $p$ reads \citep{h2005eu2}:

\begin{equation}
V_p=V_{p1}+V_{p2}
\label{equ:Vpert}
\end{equation}
with

\begin{equation}
V_{p1}=-\frac{3}{2}C\frac{\mathcal{G}M_p}{d_p^3}\big[\gamma_1(x_p^2+y_p^2)+\gamma_2(x_p^2-y_p^2)\big]
\label{equ:Vpert1}
\end{equation}
and

\begin{eqnarray}
V_{p2} & = & -\frac{15}{4}CJ_{2p}\frac{\mathcal{G}M_p}{d_p^3}\Big(\frac{\mathcal{R}_p}{d_p}\Big)^2 \nonumber \\
& \times & \big[\gamma_1(x_p^2+y_p^2)+\gamma_2(x_p^2-y_p^2)\big],
\label{equ:Vpert2}
\end{eqnarray}
where $\mathcal{G}$ is the gravitational constant, $M_p$ the mass of the perturber, $J_{2p}$ its $J_2$, $R_p$ its mean radius, $d_p$ the distance between the perturber's and Janus (or Epimetheus)' centers of mass, and $x_p$ and $y_p$ the two first components of the unit vector pointing to the center of mass of the perturber, from the center of mass of the body, in the reference frame $(\vec{f_1},\vec{f_2},\vec{f_3})$. $V_{p1}$ expresses the perturbation due to a pointmass perturber, while $V_{p2}$ represents the perturbation due to its $J_2$, assuming that the body is in the equatorial plane of the perturber. As shown in \citep{h2005eu2}, it is a good approximation if the sine of the angle between Saturn's equatorial plane and the orbit is small. Since this quantity is always smaller than $10^{-2}$ (Tab.\ref{tab:zetajan} and \ref{tab:zetaepim}), we can consider this approximation as satisfying. This assertion assumes that the obliquity is very small, what will be checked in this study. 

\par Usually the orbital ephemerides give us the location of the perturber in the inertial frame, so we have to perform 5 rotations to convert the coordinates from the inertial frame to $(\vec{f_1},\vec{f_2},\vec{f_3})$. More precisely, if we name $(x_i,y_i,z_i)^T$ the unit vector locating the perturber's center of mass in the inertial frame, we have

\begin{equation}
\left(\begin{array}{c}
x_p \\
y_p \\
z_p
\end{array}\right)
=R_3(-l)R_1(-J)R_3(-g)R_1(-K)R_3(-h)\left(\begin{array}{c}
x_i \\
y_i \\
z_i
\end{array}\right)
\label{equ:passage}
\end{equation}
with

\begin{equation}
R_3(\phi)=\left(\begin{array}{ccc}
\cos\phi & -\sin\phi & 0 \\
\sin\phi & \cos\phi & 0 \\
0 & 0 & 1
\end{array}\right)
\label{equ:r3}
\end{equation}
and

\begin{equation}
R_1(\phi)=\left(\begin{array}{ccc}
1 & 0 & 0 \\
0 & \cos\phi & -\sin\phi \\
0 & \sin\phi & \cos\phi 
\end{array}\right).
\label{equ:r1}
\end{equation}

\par Finally, the total Hamiltonian of the problem reads:

\begin{eqnarray}
H & = & \frac{nP^2}{2}+\frac{n}{8}\big[4P-\xi_q^2-\eta_q^2\big] \nonumber \\
 & & \times\Big[\frac{\gamma_1+\gamma_2}{1-\gamma_1-\gamma_2}\xi_q^2+\frac{\gamma_1-\gamma_2}{1-\gamma_1+\gamma_2}\eta_q^2\Big] \nonumber \\
 & & -\frac{3}{2n}\frac{\mathcal{G}M_{\saturn}}{d_{\saturn}^3}\Big(1+\frac{5}{2}J_{2{\saturn}}\Big(\frac{\mathcal{R}_{\saturn}}{d_{\saturn}}\Big)^2\Big) \nonumber \\
 & & \times  \big[\gamma_1(x_{\saturn}^2+y_{\saturn}^2)+\gamma_2(x_{\saturn}^2-y_{\saturn}^2)\big],
\label{equ:Htotal}
\end{eqnarray}
where the index $\saturn$ stands for Saturn. We will use this Hamiltonian for a numerical study of the rotation. An analytical study can show that the Hamiltonian (\ref{equ:Htotal}) can be reduced to 

\begin{equation}
\mathcal{H}(u,v,w,U,V,W)=\omega_uU+\omega_vV+\omega_wW+\mathcal{P}(u,v,w,U,V,W)
\label{equ:quadra}
\end{equation}
where $\mathcal{P}$ represents a perturbation, and the three constants $\omega_u$, $\omega_v$ and $\omega_w$ are the periods of the free oscillations around the equilibrium defined by the Cassini Laws. This last Hamiltonian is obtained after several canonical transformations, the first one consisting in expressing the resonant arguments $\sigma=p-\lambda+\pi$ and $\rho=r+\ascnode$ respectively associated with the $1:1$ spin-orbit resonance and with the orientation of the angular momentum, $\lambda$ and $\ascnode$ being the orbital variables defined above. The complete calculation is beyond the scope of this paper, the reader can find details in \citep{h2005io,h2005eu,nlv2008}.

\par Tab.\ref{tab:param} gives the values of the physical parameters we used in our numerical integrations. They come mainly from spacecraft datas, Pioneer \& Voyager for Saturn, and Cassini for Janus and Epimetheus.


\begin{table}
\begin{center}
\caption[Physical parameters of Saturn, Janus and Epimetheus]{Physical parameters of Saturn, Janus and Epimetheus. The gravity parameters $J_2$ and $C_{22}$ of Janus and Epimetheus have been derived from the moments of inertia given by \citet{ttb2009}. \label{tab:param}}
\begin{tabular}{l|ll}
\tableline\tableline
Parameters & Values & References \\
\tableline
Saturn & & \\
\tableline
$\mathcal{R_{\saturn}}$ (mean) & $58232$ km & IAU 2006 \citep{saacch2007} \\
(equatorial) & $60330$ km & Pioneer \& Voyager  \citep{ca1989} \\
$J_{2\saturn}$ & $1.6298\times10^{-2}$ & Pioneer \& Voyager \citep{ca1989} \\
$\mathcal{G}M_{\saturn}$ & $3.7931272\times10^7$ km$^3$.s$^{-2}$ & Pioneer \& Voyager \citep{ca1989} \\
\tableline
Janus & & \\
\tableline
$\mathcal{G}M_J$ & $0.12660$ km$^3$.s$^{-2}$ & \citep{jspbcem2008} \\
$\mathcal{R}$ & $89.5$ km & Cassini \citep{ttb2009} \\
$A/M\mathcal{R}^2$ & $0.360$ & Cassini \citep{ttb2009} \\
$B/M\mathcal{R}^2$ & $0.407$ & Cassini \citep{ttb2009} \\
$C/M\mathcal{R}^2$ & $0.470$ & Cassini \citep{ttb2009} \\
$J_2$ & $8.65\times10^{-2}$ & - \\
$C_{22}$ & $1.175\times10^{-2}$ & - \\
\tableline
Epimetheus & & \\
\tableline
$\mathcal{G}M_E$ & $0.03513$ km$^3$.s$^{-2}$ & \citep{jspbcem2008} \\
$R$ & $58.1$ km & Cassini \citep{ttb2009} \\
$A/M\mathcal{R}^2$ & $0.328$ & Cassini \citep{ttb2009} \\
$B/M\mathcal{R}^2$ & $0.469$ & Cassini \citep{ttb2009} \\
$C/M\mathcal{R}^2$ & $0.476$ & Cassini \citep{ttb2009} \\
$J_2$ & $7.75\times10^{-2}$ & - \\
$C_{22}$ & $3.525\times10^{-2}$ & - \\
\tableline
\end{tabular}
\end{center}
\end{table}

\subsection{Numerical integrations}

\par In order to integrate numerically the system, we first express the coordinates of the perturber $(x_{\saturn},y_{\saturn})$ with the numerical ephemerides and the rotations given in (Eq.\ref{equ:passage}), in the body frame $(\vec{f_1},\vec{f_2},\vec{f_3})$. As explained before, the ephemerides have been obtained in interpolating the HORIZONS cartesian coordinates of Janus and Epimetheus with cubic splines. This way, we get coordinates depending of the canonical variables. Then we derive the equations coming from the Hamiltonian (\ref{equ:Htotal}):

\begin{eqnarray}
\frac{dp}{dt} =  \frac{\partial H}{\partial P}, & &  \frac{dP}{dt} = -\frac{\partial H}{\partial p},  \nonumber \\
\frac{dr}{dt} = \frac{\partial H}{\partial R}, & & \frac{dR}{dt}=-\frac{\partial H}{\partial r}, \nonumber \\
\frac{d\xi_q}{dt} = \frac{\partial H}{\partial \eta_q}, & & \frac{d\eta_q}{dt}=-\frac{\partial H}{\partial \xi_q}. \label{equ:equhamil}
\end{eqnarray}

\par We integrated over 100 years using the Adams-Bashforth-Moulton 10th order predictor-corrector integrator. The solutions consist of two parts, the forced one, directly due to the perturbation, and the free one, that depends on the initial conditions. The initial conditions should be as close as possible to the exact equilibrium, to have low amplitudes of the free librations. Several methods exist to reduce the amplitudes of the free librations:

\begin{itemize}

\item \citet{br2007} propose to fit the mean initial conditions in order to locate the spin-orbit system at its center of libration,

\item \citet{pym2007} add a damping in the equations that reduces the amplitude of the free librations,

\item \citet{ym2006}, in the framework of a numerical integration of the spin and of the orbit of Mercury, start from a simple Sun-Mercury system in which the equilibrium is very easy to determine analytically, and slowly switch on the planetary perturbations in order to create an adiabatic deviation of the equilibrium without creation of any free libration,

\item \citet{n2009} use an iterative scheme to remove the free librations from the initial conditions. This last approach is quite similar to the one of \citet{lg2000} in the framework of a computer-assisted proof of the KAM theorem.

\end{itemize}

\par The method we used here is inspired from the one of \citet{pym2007}. In fact, we performed a first numerical integration with a damping, then a reverse integration without any damping. This way, we got initial conditions that we use to get the solution of the rigid rotation of Janus and Epimetheus without any free rotational energy. We finally use these last initial conditions to get a rotational state in which the energy is minimized. We also avoid any shift of the equilibrium that a too fast dissipation might induce. The reasons why we want to minimize the free component of the solutions is first because this solution is expected to have been damped by dissipations, and second because a signal from which useless components have been dropped will be more efficiently analyzed.

\par Our model of dissipation is just a mathematical one, i.e. without considering the physical cause of the dissipation, because it just aimed at finding initial condition and not at studying the dissipation itself. We added to every equation a dissipative term alike $-\alpha(x-x^*)$ where $x^*$ is an approximation of the equilibrium of the variable $x$, and $\alpha$ a positive constant. Our only constraint on the value of $\alpha$ is that the dissipation is adiabatic, because a too fast dissipation changes the equilibrium of the system (see for instance the influence of tides on the equilibrium state of Venus in \citep{cln2003,cl2003}, respectively for the theory and the numerical confirmation).

\subsection{The output variables}

\par In order to deliver theories of rotation that can be easily compared with observations, we chose to express our results in the following variables:

\begin{itemize}

\item Longitudinal librations,

\item Latitudinal librations,

\item Orbital obliquity $\epsilon$ (the orientation of the angular momentum of Janus/Epimetheus with respect to the normal to the instantaneous orbital plane),

\item Motion of the rotation axis about the pole axis.

\end{itemize}

\par There are at least two ways to define the longitudinal librations. We can for instance consider the librations about the exact synchronous rotation, i.e. $p-<n>t$, usually called \emph{physical librations}. The determination of a "mean" mean motion for Janus and Epimetheus is far to be obvious because of the orbital swaps. We can either consider a mean motion averaged over several swaps, and get averaged librations that cannot actually be observed at a given date, or consider two different mean motions for each satellite, in assuming that the mean motion is constant between two swaps. In order to keep this assumption valid, we shall use two intervals of study, necessarily smaller than 4 years, that are far enough from the swaps to not be affected by the transitions. We determined the numerical values of the mean motions out of the swaps in using the slope of the mean longitudes of each bodies over the time spans [1998.5:2001.7] and [2002.5:2005.7] (cf. Tab.\ref{tab:meanmotion}).


\begin{table}[ht]
\centering
\caption[Mean motions of Janus and Epimetheus]{Mean motions of Janus and Epimetheus between two orbital swaps, determined in using the JPL HORIZONS ephemerides. These values for the period $[2002.5:2005.7]$ are close to the mean motions given by \citet{jspbcem2008}, fitted on quite the same time span.\label{tab:meanmotion}}
\begin{tabular}{c|cc}
\tableline\tableline
 & $[1998.5:2001.7]$ & $[2002.5:2005.7]$ \\
\tableline
Janus & $3304.35631\pm5.658\times10^{-5}$ & $3303.67315\pm5.662\times10^{-5}$ \\
Epimetheus & $3302.78370\pm1.259\times10^{-3}$ & $3305.24602\pm1.255\times10^{-3}$ \\
\tableline
\end{tabular}
\end{table}

\par Another way to consider the longitudinal librations is to work on the librations about the Janus-Saturn (or Epimetheus-Saturn) direction. We will call these librations \emph{tidal librations} because they represent the misalignment of the tidal bulge of the satellite. The reader can find graphical descriptions of these librations in \citet{md1999}, Fig.5.16 or in \citet{ttb2009}, Fig.3, where the physical librations are written as $\gamma$ and the tidal librations as $\psi$. The difference betwteen the two is the \emph{optical libration} ($\psi-\gamma=2e\cos nt$), which arises from Kepler's Third Law. There is a mistake in the Fig. 5.16 of \citet{md1999}, that illustrates a dynamically forbidden state (the satellite's long axis can never point between the direction towards Saturn an dthe direction towards the empty focus), the equations being correct. This problem is corrected in \citet{ttb2009}.

\par The latitudinal librations are the North-South librations of the large axis of the considered body in the saturnocentric reference frame that follows the orbital motion of the body. They are analogous to the tidal librations that are the East-West librations. In order to get to the tidal longitudinal librations and the latitudinal librations, we first should express the unit vector $\vec{f_1}$ (i.e. the direction of Janus/Epimetheus' long axis) in the inertial frame $(\vec{e_1},\vec{e_2},\vec{e_3})$. From (Eq.\ref{equ:passage}) and the definitions of the Andoyer modified variables (Eq.\ref{equ:modified}), we get:

\begin{eqnarray}
\vec{f_1} & = & (\cos r (\cos(p+r-l)\cos l-\sin(p+r-l)\cos J\sin l)+ \nonumber \\
 & & \sin r(\cos K(\sin(p+r-l)\cos l+\cos (p+r-l)\cos J\sin l)-\sin K\sin J\sin l)) \vec{e_1} \nonumber \\
 & + & (-\sin r (\cos(p+r-l)\cos l-\sin(p+r-l)\cos J\sin l)+ \nonumber \\
 & & \cos r(\cos K(\sin(p+r-l)\cos l+\cos (p+r-l)\cos J\sin l)-\sin K\sin J\sin l)) \vec{e_2} \nonumber \\
 & + & (\sin K (\sin(p+r-l)\cos l+\cos(p+r-l)\cos J\sin l)+\cos K\sin J\sin l) \vec{e_3}. \label{equ:f1}
\end{eqnarray}
The tidal longitudinal librations $\psi$ and the latitudinal ones $\eta$ are found this way:

\begin{equation}
\label{equ:psicross}
\psi=\vec{t}\cdot\vec{f_1}
\end{equation}
and

\begin{equation}
\label{equ:etacross}
\eta=\vec{n}\cdot\vec{f_1},
\end{equation}
where $\vec{n}$ is the unit vector normal to the orbit plane, and $\vec{t}$ the tangent to the trajectory. We get these last two vectors by:

\begin{equation}
\label{equ:vecnorm}
\vec{n}=\frac{\vec{x}\times\vec{v}}{||\vec{x}\times\vec{v}||}
\end{equation}
and

\begin{equation}
\label{equ:vectang}
\vec{t}=\frac{\vec{n}\times\vec{x}}{||\vec{n}\times\vec{x}||},
\end{equation}
where $\vec{x}$ is the position vector of the body, and $\vec{v}$ its velocity.

\par Finally, the motion of the rotation axis about the pole is derived from the wobble $J$, it is given by the two variables $Q_1$ and $Q_2$ defined as:

\begin{equation}
\label{equ:Q1}
Q_1=\sin J \sin l \bigg(1+\frac{J_2+2C_{22}}{C}\bigg)
\end{equation}
and

\begin{equation}
\label{equ:Q2}
Q_2=\sin J \cos l \bigg(1+\frac{J_2-2C_{22}}{C}\bigg),
\end{equation}
they are the first two components of the unit vector pointing at the instantaneous North Pole of Janus' (respectively Epimetheus') rotation axis, in the body frame of Janus (or Epimetheus). These quantities are finally multiplied by the polar radius of the satellite ($76.3$ km for Janus and $53.1$ km for Epimetheus \citep{ttb2009}) to get a deviation in meters.

\par As for the previous study and for the orbital ephemerides, we will give these solutions under a semi-analytical (or synthetic) form, thanks to the frequency analysis.

\subsection{The results}

\par As said above, there are two parts in the solutions: the free and the forced ones. The free solutions depend only on the initial conditions and are assumed to be damped because of dissipations (in particular tidal dissipations). It is anyway interesting to study it, at least to compute the three periods associated, in case of one of them would be close to a resonance with a forced contribution. We detect the free terms thanks to the frequency analysis and their values are gathered in Tab.\ref{tab:freqlib}.


\begin{table}[ht]
 \centering
\caption{Periods of the free librations, determined numerically.\label{tab:freqlib}}
\begin{tabular}{l|cc}
\tableline\tableline
 & Janus & Epimetheus \\
\tableline
$T_u$ & $1.26713$ d & $0.74717$ d \\
$T_v$ & $2.17884$ d & $1.80386$ d \\
$T_w$ & $2.09798$ d & $5.54234$ d \\
\tableline
\end{tabular}
\end{table}

\par The forced solutions are gathered in the Tab.\ref{tab:longijan} to \ref{tab:obli2}. In all the following tables, the identification of the periodic contribution has been made in checking the frequencies and the phases. In particular, the phases, given at J2000, are very useful to discriminate 2 contributions with very close frequencies, like $\varpi_J$ and $\varpi_E$. We here present all the sinusoidal terms actually detected by the frequency analysis algorithm, except when we precise in the caption that a cut-off has been made.



\begin{table}[ht]
 \centering
\caption[Tidal longitudinal librations of Janus]{Tidal longitudinal librations $\psi$ of Janus. The phases are given at J2000, the series are in sine.\label{tab:longijan}}
\begin{tabular}{r|rrrrrr}
\tableline\tableline
 N & $\lambda$ & $\phi$ & $\varpi_J$ & Amplitude & Phase & Period \\
\tableline
 $1$ & $1$ & -    & $-1$ &  $1.03^{\circ}$ & $116.360^{\circ}$ & $0.69735$ d \\
 $2$ & $1$ & $-1$ & $-1$ &  $17.54$ arcmin & $-61.337^{\circ}$ & $0.69752$ d \\
 $3$ & $1$ &  $1$ & $-1$ & $-17.52$ arcmin & $114.056^{\circ}$ & $0.69719$ d \\
 $4$ & $1$ &  $2$ & $-1$ &   $3.04$ arcmin & $112.262^{\circ}$ & $0.69702$ d \\
 $5$ & $1$ & $-2$ & $-1$ &   $2.99$ arcmin & $120.439^{\circ}$ & $0.69769$ d \\
 $6$ & $1$ & $-3$ & $-1$ &  $-1.70$ arcmin & $123.544^{\circ}$ & $0.69785$ d \\
 $7$ & $1$ &  $3$ & $-1$ &   $1.67$ arcmin & $-70.825^{\circ}$ & $0.69685$ d \\
 \tableline
\end{tabular}
\end{table}

\begin{table}[ht]
 \centering
\caption[Latitudinal librations of Janus]{Latitudinal librations $\eta$ of Janus, in arcseconds. The series are in sine.\label{tab:latijan}}
\begin{tabular}{r|rrrrrr}
\tableline\tableline
 N & $\lambda$ & $\phi$ & $\ascnode_J$ & Amplitude & Phase & Period \\
\tableline
 $1$ & $1$ & -    & $-1$ & $-6.18$ & $-48.717^{\circ}$ & $0.69186$ d \\
 $2$ & $1$ &  $1$ & $-1$ &  $1.79$ & $-50.972^{\circ}$ & $0.69170$ d \\
 $3$ & $1$ & $-1$ & $-1$ & $-1.79$ & $113.539^{\circ}$ & $0.69202$ d \\
 $4$ & $1$ &  $2$ & $-1$ & $-0.31$ & $-53.312^{\circ}$ & $0.69153$ d \\
 $5$ & $1$ & $-2$ & $-1$ & $-0.31$ & $-44.149^{\circ}$ & $0.69219$ d \\
 \tableline
\end{tabular}
\end{table}

\par For Janus, the librations are dominated by short-period contributions essentially due to the mean anomaly $\lambda-\varpi_J$ for the physical longitudinal librations $\gamma$, and to $\lambda-\ascnode_J$ for the latitudinal ones. The physical cause of these librations is the variations of the distance Saturn-Janus because of the eccentricity. We can also notice the other contributions, with this form for latitudinal librations:

\begin{equation}
A\big(\sin(\lambda+i\phi-\ascnode_J)+\sin(\lambda-i\phi-\ascnode_J)\big)=2A\sin(\lambda-\ascnode_J)\cos(i\phi)
\label{equ:battements2}
\end{equation}
where $i$ is an integer. For longitudinal librations, $\varpi_J$ should replace $\ascnode_J$ in (Eq.\ref{equ:battements2}).  We here consider that the amplitudes associated are the same, in fact they have relative differences smaller that $10\%$. Moreover, the phases are consistent with the frequencies, because they are contained in the proper modes $\lambda$, $\phi$ and $\ascnode$. If the signs of the amplitudes are opposite, the right-hand side of Eq.(\ref{equ:battements2}) is $2A\cos(\lambda-\ascnode_J)\sin(i\phi)$, so the resulting wave has the same visual aspect. These contributions result in 0.697-d-periodic beatings in $4/i$-y-periodic envelopes (where i=1,2,3,\ldots , so periods of 4, 2, 1.33, \ldots years, with decreasing amplitudes associated), they are librations due to the orbital swaps. We can see that for Janus, they remain small compared to the "classical" librations (i.e. just involving $\lambda$ and the node/pericentre). 

\par The Tab.\ref{tab:longiepi} and \ref{tab:latiepi} present the same variables for Epimetheus. The striking difference is that the effects of the orbital swaps dominate the dynamics, whereas the contributions $\lambda-\varpi_E$ and $\lambda-\ascnode_E$ are of smaller amplitude. This difference is due to the amplitude of the orbital swap for Epimetheus, that is larger than for Janus because of the mass ratios of the two satellites, as we already noticed.



\begin{table}[ht]
 \centering
\caption[Tidal longitudinal librations of Epimetheus]{Tidal longitudinal librations $\psi$ of Epimetheus. The series are in sine.\label{tab:longiepi}}
\begin{tabular}{r|rrrrrr}
\tableline\tableline
 N & $\lambda$ & $\phi$ & $\varpi_E$ & Amplitude & Phase & Period \\
\tableline
 $1$ & $1$ & $-1$ & $-1$ &   $5.19^{\circ}$ & $-170.621^{\circ}$ & $0.69752$ d \\
 $2$ & $1$ &  $1$ & $-1$ &  $-5.17^{\circ}$ &    $4.964^{\circ}$ & $0.69719$ d \\
 $3$ & $1$ &  $2$ & $-1$ &  $-3.93^{\circ}$ & $-177.326^{\circ}$ & $0.69702$ d \\
 $4$ & $1$ & $-2$ & $-1$ &  $-3.92^{\circ}$ & $-168.330^{\circ}$ & $0.69769$ d \\
 $5$ & $1$ & -    & $-1$ &  $-2.58^{\circ}$ & $-172.826^{\circ}$ & $0.69735$ d \\
 $6$ & $1$ &  $3$ & $-1$ &  $-1.09^{\circ}$ &    $0.215^{\circ}$ & $0.69685$ d \\
 $7$ & $1$ & $-3$ & $-1$ &   $1.07^{\circ}$ & $-165.862^{\circ}$ & $0.69785$ d \\
 $8$ & $1$ & $-4$ & $-1$ &  $28.14$ arcmin  &   $15.891^{\circ}$ & $0.69802$ d \\
 $9$ & $1$ &  $4$ & $-1$ &  $27.50$ arcmin  &   $-1.553^{\circ}$ & $0.69669$ d \\
$10$ & $1$ &  $5$ & $-1$ &  $20.03$ arcmin  &  $175.657^{\circ}$ & $0.69652$ d \\
$11$ & $1$ & $-5$ & $-1$ & $-19.56$ arcmin  &   $18.683^{\circ}$ & $0.69819$ d \\
\tableline
\end{tabular}
\end{table}

\begin{table}[ht]
 \centering
\caption[Latitudinal librations of Epimetheus]{Latitudinal librations $\eta$ of Epimetheus, in arcseconds. The series are in sine.\label{tab:latiepi}}
\begin{tabular}{r|rrrrrr}
\tableline\tableline
 N & $\lambda$ & $\phi$ & $\ascnode_E$ & Amplitude & Phase & Period \\
\tableline
 $1$ & $1$ &  $1$ & $-1$ &  $5.75$ & $-89.666^{\circ}$ & $0.69170$ d \\
 $2$ & $1$ & $-1$ & $-1$ & $-5.75$ &  $94.847^{\circ}$ & $0.69202$ d \\
 $3$ & $1$ &  $2$ & $-1$ &  $4.49$ &  $88.082^{\circ}$ & $0.69153$ d \\
 $4$ & $1$ & $-2$ & $-1$ &  $4.48$ &  $97.010^{\circ}$ & $0.69219$ d \\
 $5$ & $1$ & -    & $-1$ &  $2.68$ &  $92.593^{\circ}$ & $0.69186$ d \\
 $6$ & $1$ &  $3$ & $-1$ &  $1.30$ & $-94.153^{\circ}$ & $0.69137$ d \\
 $7$ & $1$ & $-3$ & $-1$ & $-1.30$ &  $99.340^{\circ}$ & $0.69235$ d \\
 \tableline
\end{tabular}
\end{table}

\par  The results related to the orbital obliquity $\epsilon$ are gathered in Tab.\ref{tab:obli2}, for the two satellites. The phases are not given, but the reader can guess them from Tab.\ref{tab:propmod} and the signs of the amplitude (a minus sign means a phase shift of $180^{\circ}$). The most interesting is the first line, giving the mean theoretical obliquities, respectively 5.95 and 10.8 arcseconds. Their variations, smaller than one arcsec, are too small to be detected from observations. As for the librations, the orbital swaps have a bigger effect on Epimetheus than on Janus, but even for Epimetheus the obliquity only varies by a few percent.


\begin{table}[ht]
 \centering
\caption[Orbital obliquity]{Orbital obliquity $\epsilon$ of the two bodies, in arcseconds. The contributions for which no amplitude is given have not been detected by the frequency analysis. It was required to remove a slope of $-5.799\times10^{-4}$ arcsec/y, probably the signature of a long-period contribution. $\varpi$ stands for $\varpi_J$ for Janus,  $\varpi_E$ for Epimetheus. Contributions associated with the nodes $\ascnode$ have been actually detected by the frequency analysis, but with smaller amplitudes ($<0.01$ arcsec for Janus, and $<0.3$ arcsec for Epimetheus). The series are in cosine.\label{tab:obli2}}
\begin{tabular}{rrrrrr}
\tableline\tableline
 $\lambda$ & $\phi$ & $\varpi$ & Amplitude & Amplitude & Period \\
 & & & Janus & Epimetheus & \\
\tableline
 -   & -    & -    & $5.94$ & $10.83$ & $\infty$ \\
 $1$ & -    & $-1$ & $0.02$ & $-0.43$ & $0.69735$ d \\
 $1$ & $-1$ & $-1$ & $0.01$ &  $0.87$ & $0.69752$ d \\
 $1$ &  $1$ & $-1$ & $-0.01$ &  $-0.87$ & $0.69719$ d \\
 $1$ &  $2$ & $-1$ & -- & $-0.66$ & $0.69702$ d \\
 $1$ & $-2$ & $-1$ & -- & $-0.66$ & $0.69769$ d \\
 \tableline
\end{tabular}
\end{table}

\par As already pointed out by \citet{bks2003}, there a small motion of the North Pole axis about the angular momentum, but also far too small to be detected ($\approx$ 1 m for the two bodies). The tables associated are in the supplemental material.

\par
\begin{figure}[ht]
\centering
\begin{tabular}{cc}
 \includegraphics[height=5cm,width=7.8cm]{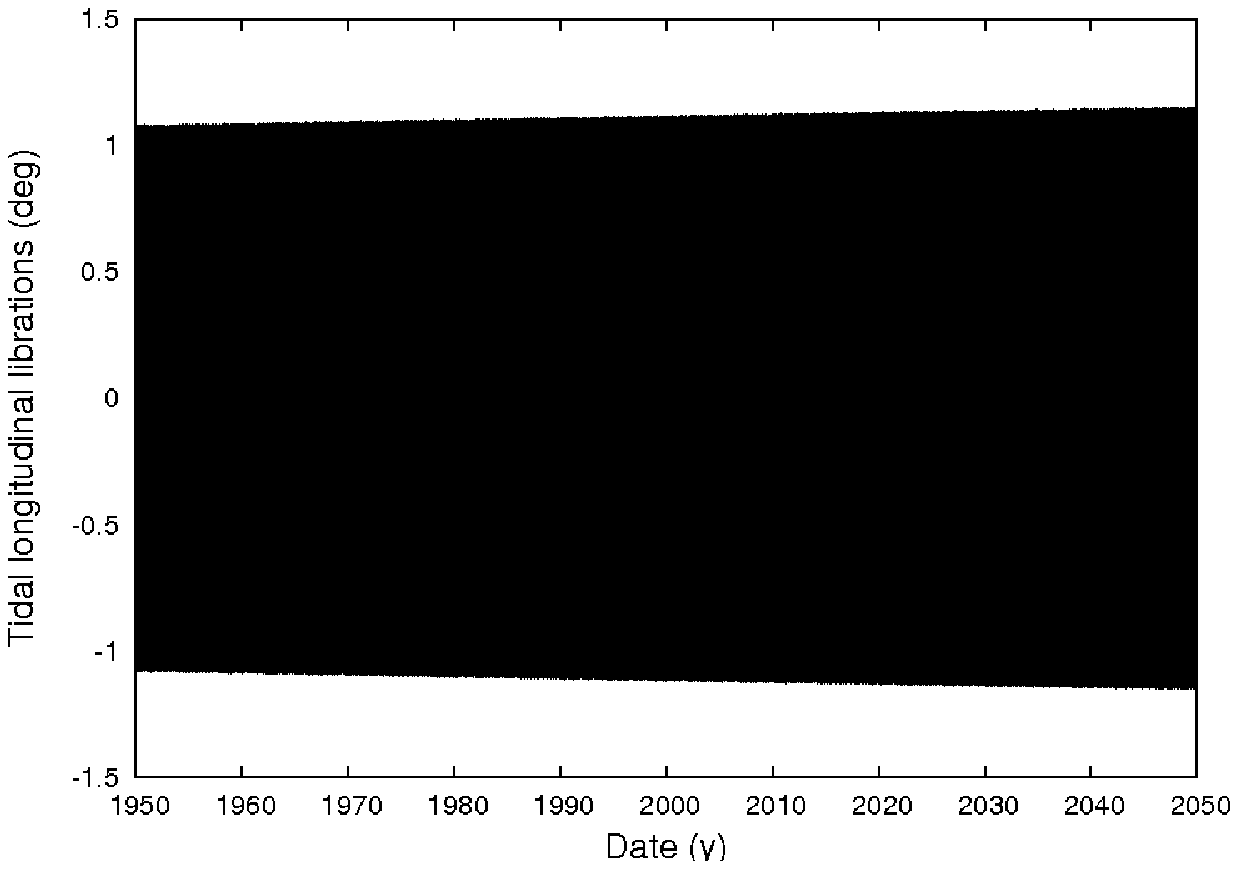} & \includegraphics[height=5cm,width=7.8cm]{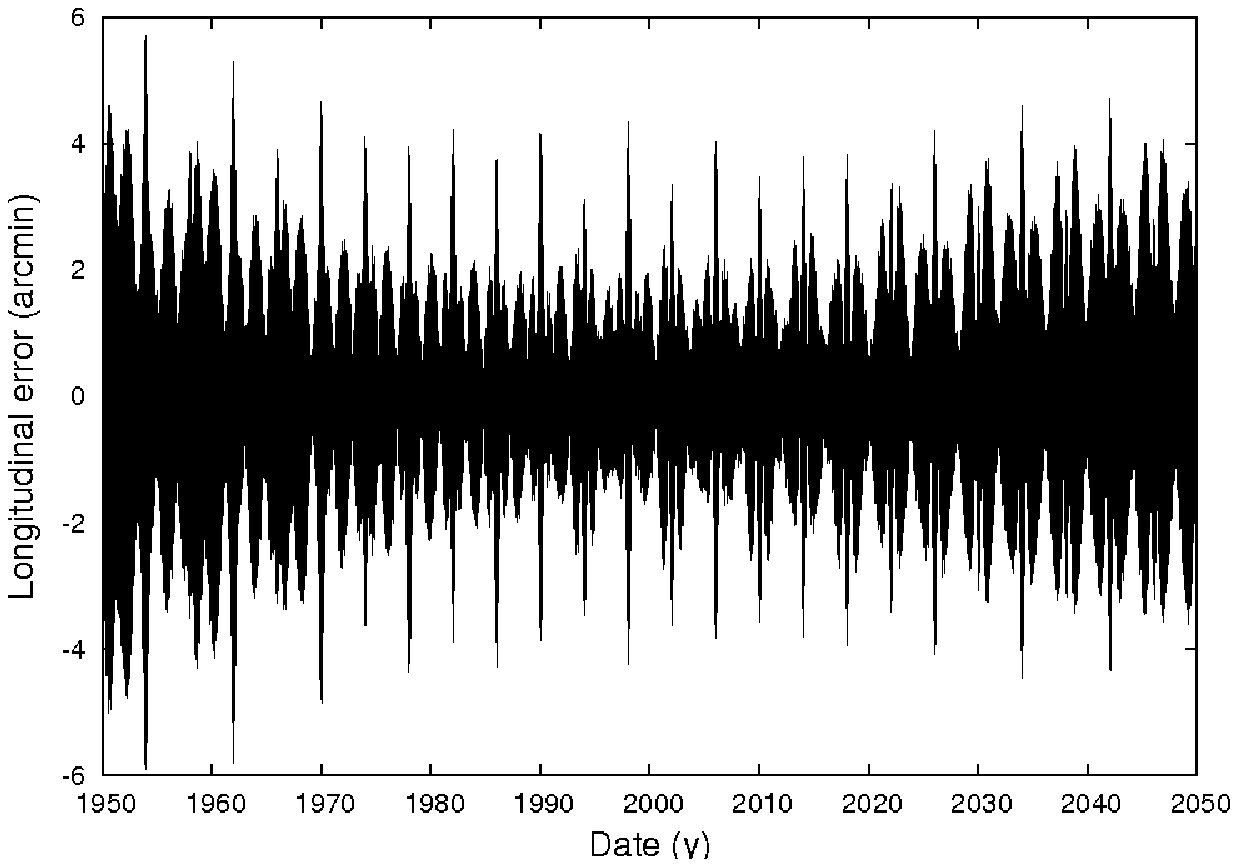} \\
(a) & (b) \\
 \includegraphics[height=5cm,width=7.8cm]{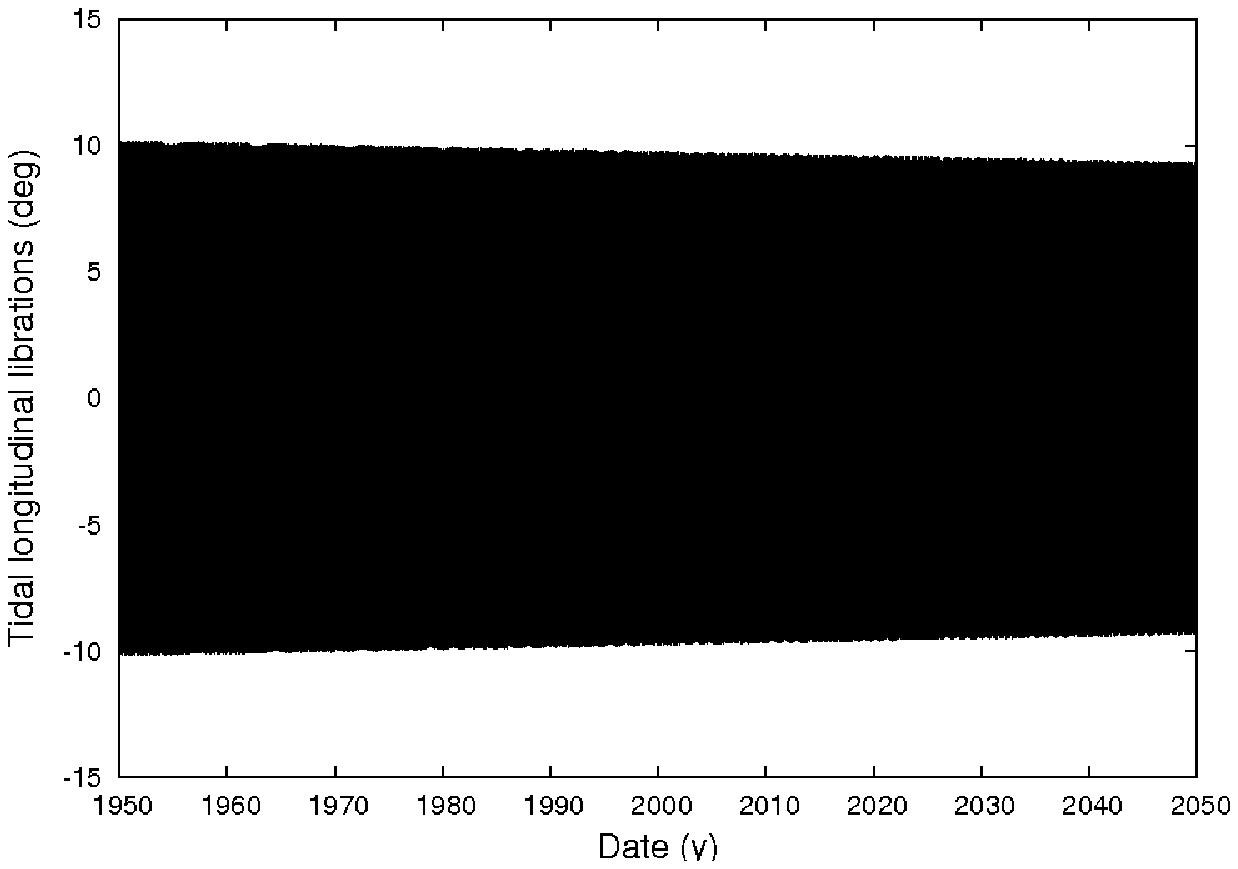} & \includegraphics[height=5cm,width=7.8cm]{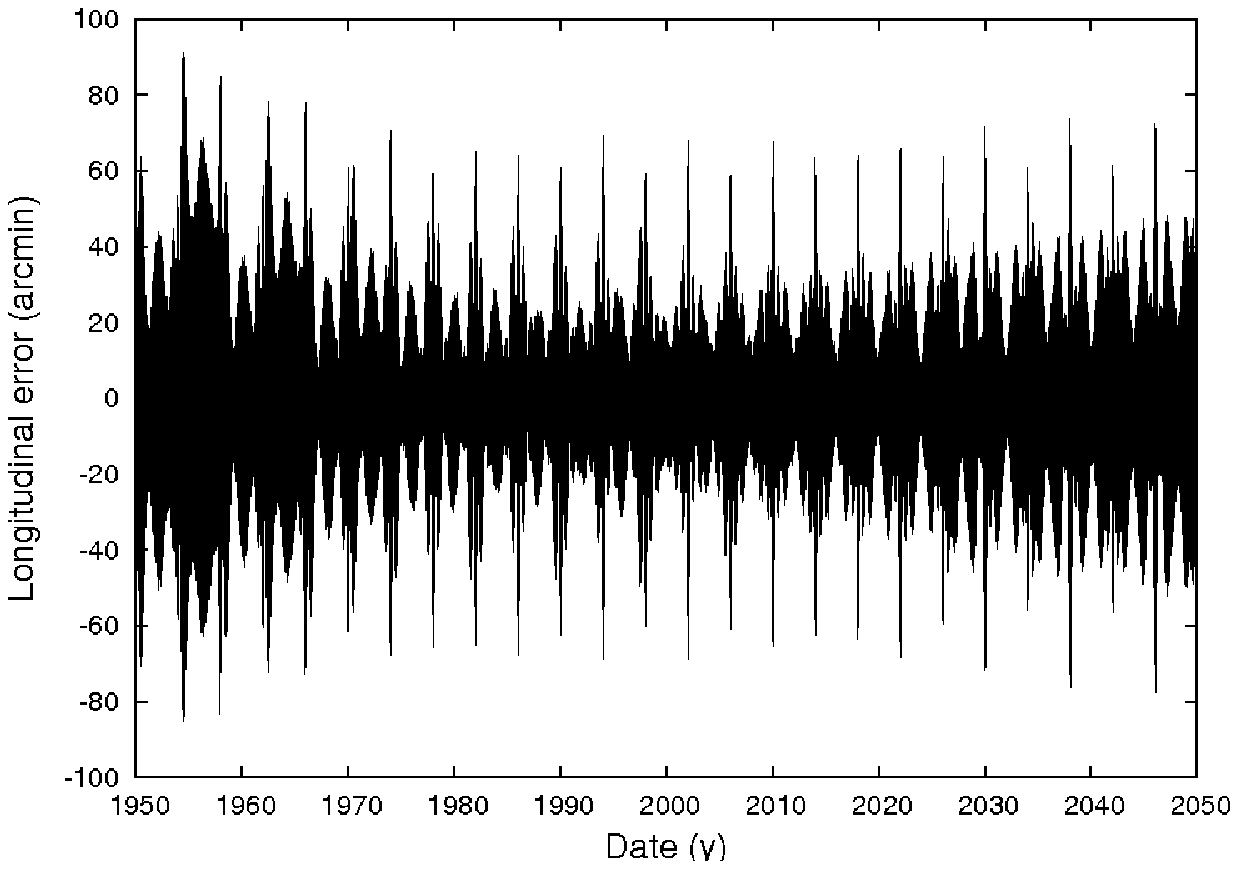} \\
(c) & (d)
\end{tabular}
\caption[Tidal longitudinal librations]{Tidal longitudinal librations $\psi$ of Janus (up) and Epimetheus (down). The panels (b) and (d) show the residuals with the quasi-periodic representations, on which we can see the signature of a 4-y periodic error, probably due to the Gibbs phenomenon at the orbital swaps.\label{fig:tidlongi}}
\end{figure}

 We now give (see Tab.\ref{tab:err}) an estimation of the error induced by the synthetic representation of the output variables. This is the maximum error over the time interval of the study (i.e. [1950:2050]), that we compare with the maximum amplitude of the variable. In comparing this maximum error with the amplitudes of the sinusoidal terms given in Tab.\ref{tab:longijan} to \ref{tab:obli2}, it seems to be too high. In fact, this maximum error is due to the Gibbs phenomenon at the orbital swaps that complicates the convergence of the periodic series, as for the orbital motion (Fig.\ref{fig:distance}). Moreover, there is also an error far from the swaps, probably due to a lack of accuracy in the determination of the short periods. The Fig.\ref{fig:tidlongi} illustrates this problem for the longitudinal librations. It also shows that the error is smaller at the center of the interval ($\approx$ 1 arcmin) than at the edges ($\approx$ 2 arcmin), probably because of long-period terms that the frequency analyses did not identify.

\placetable{tab:err}

\begin{table}[ht]
\caption[Output variables of the rotation]{Maximum amplitudes and errors of the quasiperiodic representation of the output variables of the rotation, estimated from the plots. The error is mostly due to the Gibbs phenomenon at the orbital swaps, see also Fig.\ref{fig:tidlongi}. \label{tab:err}}
\centering
\begin{tabular}{c|cc}
\tableline\tableline
Variable & Amplitude & Error \\
\tableline
 & \multicolumn{2}{c}{Tidal longitudinal librations} \\
\tableline
Janus & $\approx1.1^{\circ}$ & $\approx5$ arcmin \\
Epimetheus & $\approx10^{\circ}$ & $\approx1.5^{\circ}$ \\
\tableline
 & \multicolumn{2}{c}{Latitudinal librations} \\
\tableline
Janus & $\approx6.5$ arcsec & $\approx0.9$ arcsec \\
Epimetheus & $\approx11$ arcsec & $\approx3$ arcsec \\
\tableline
 & \multicolumn{2}{c}{Obliquity $\epsilon$} \\
\tableline
Janus & $\approx6.15$ arcsec & $\approx0.2$ arcsec \\
Epimetheus & $\approx14$ arcsec & $\approx1$ arcsec \\
\tableline
\end{tabular}
\end{table}


\par
\begin{figure}[ht]
\centering
\begin{tabular}{cc}
 \includegraphics[height=5cm,width=7.8cm]{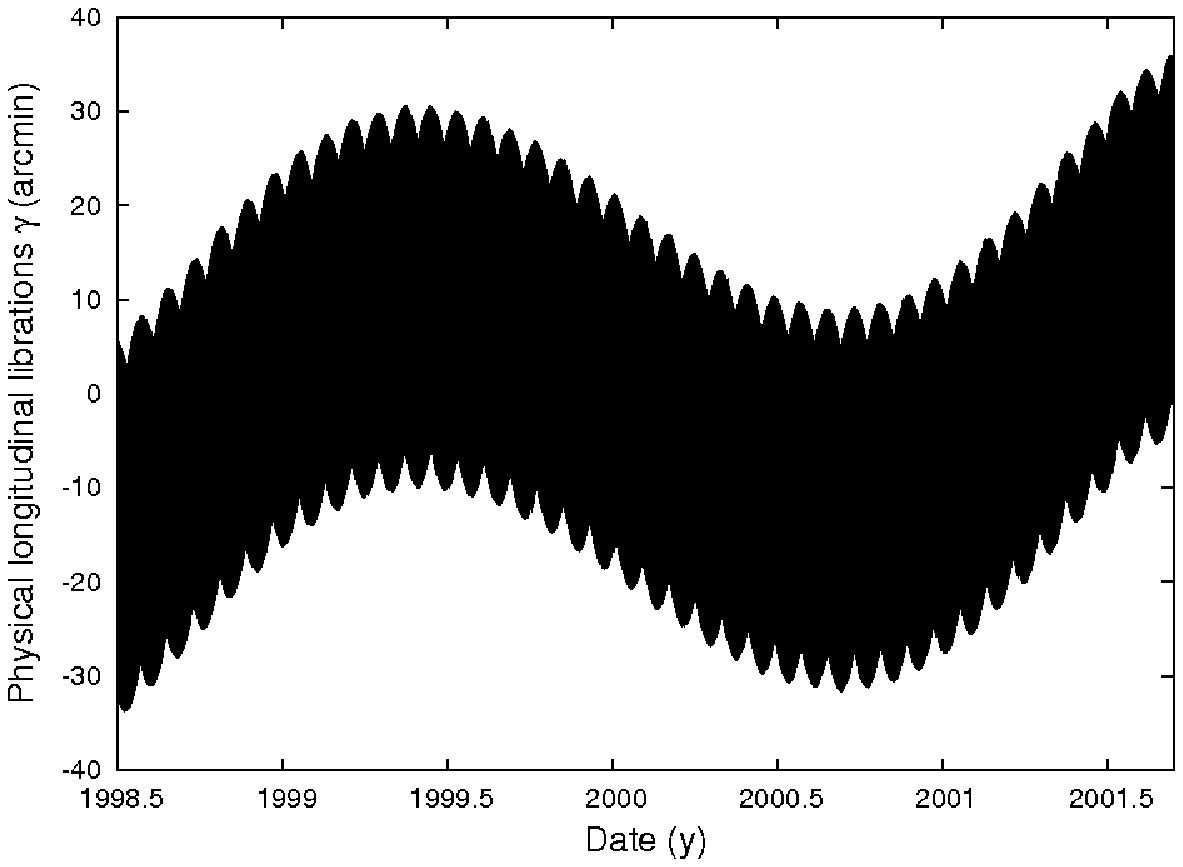} & \includegraphics[height=5cm,width=7.8cm]{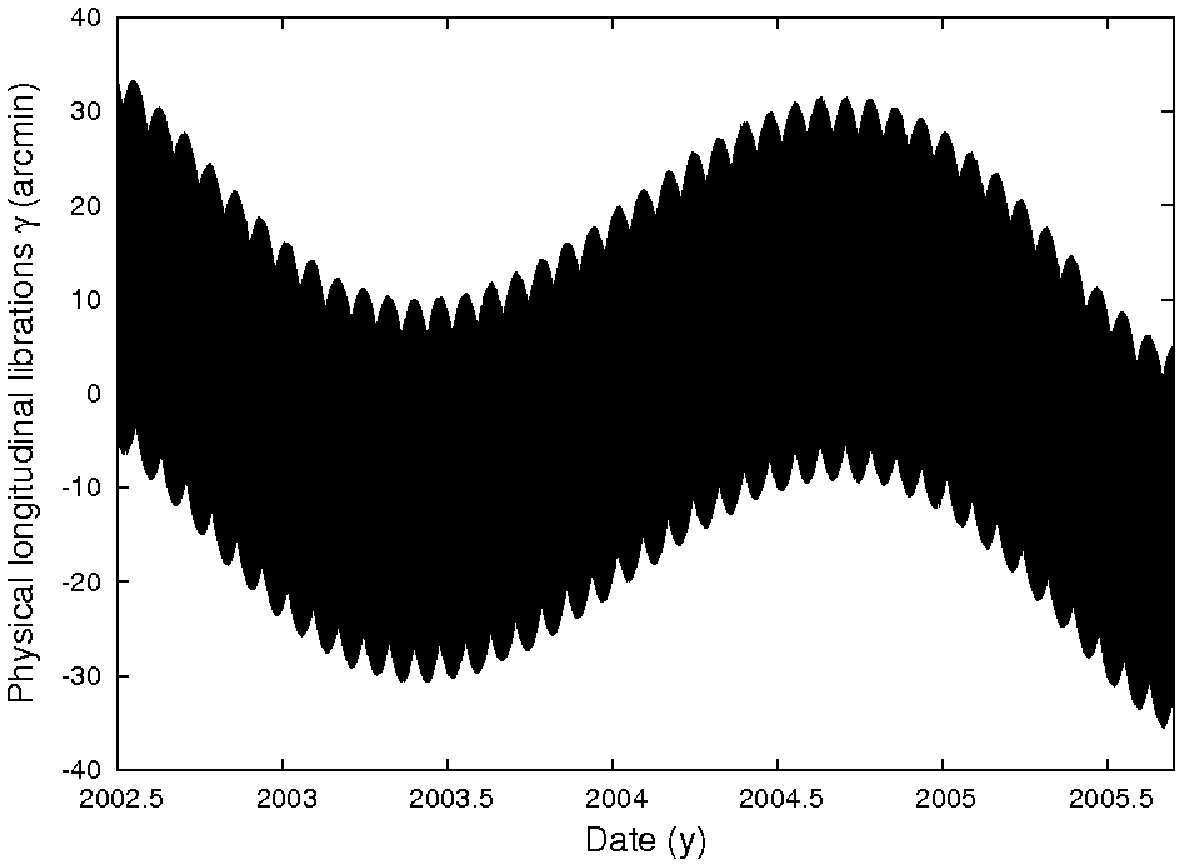} \\
(a) & (b) \\
 \includegraphics[height=5cm,width=7.8cm]{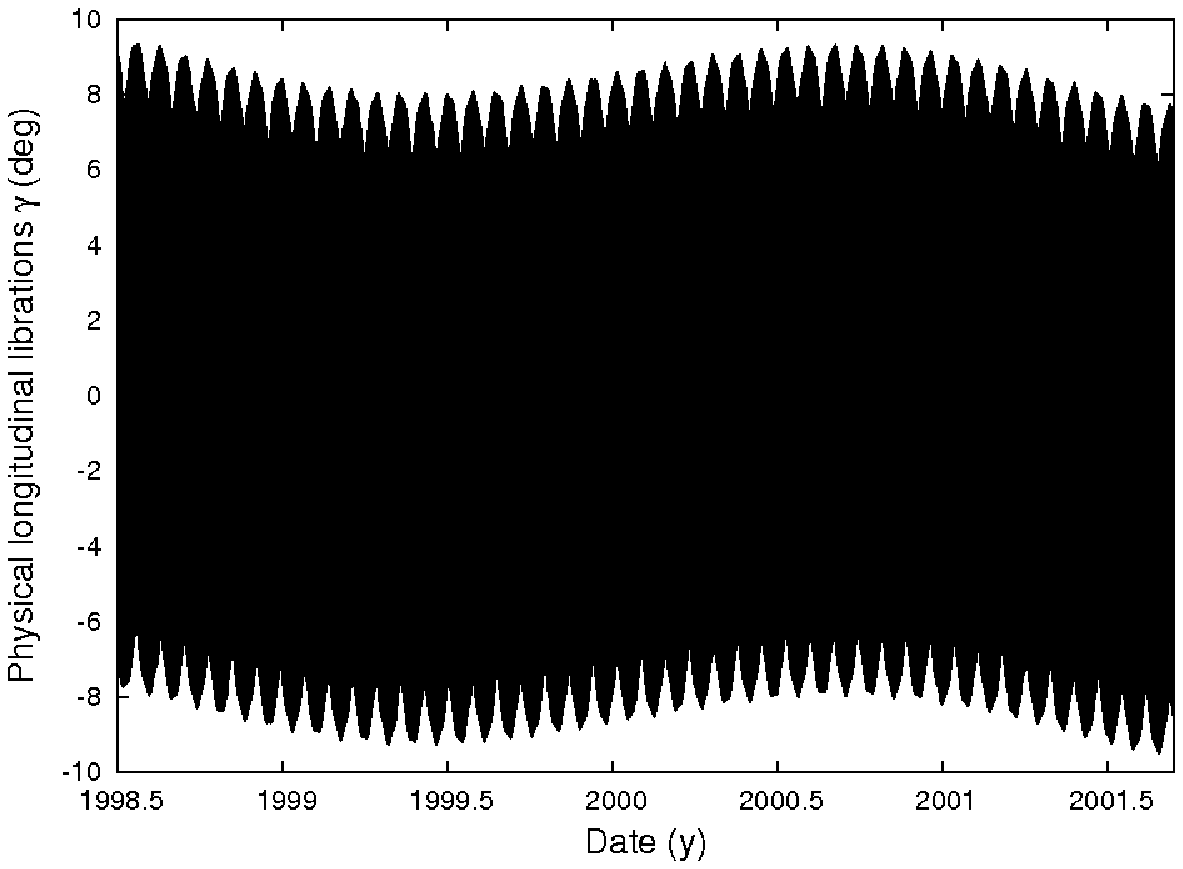} & \includegraphics[height=5cm,width=7.8cm]{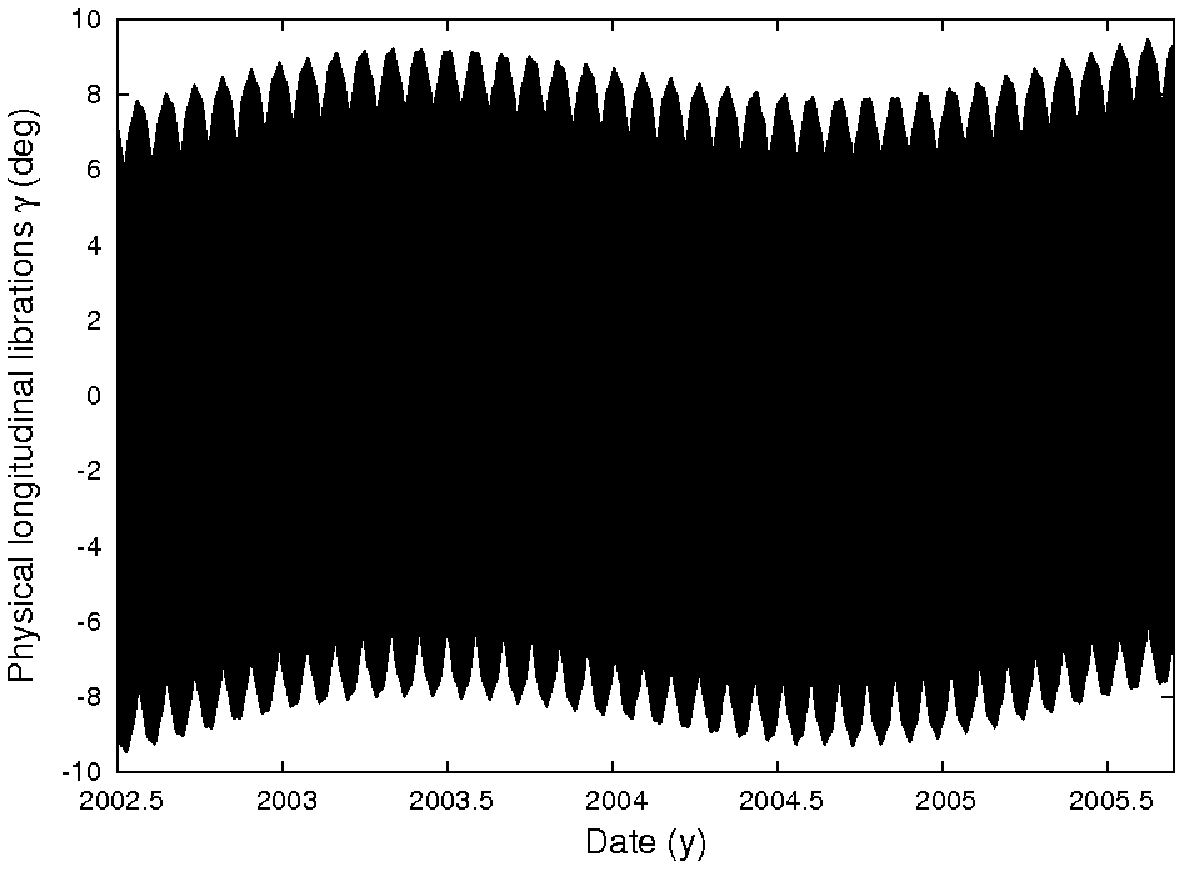} \\
(c) & (d)
\end{tabular}
\caption[Physical longitudinal librations]{Physical longitudinal librations $\gamma$ of Janus (up) and Epimetheus (down). These librations have been obtained in removing a fitted mean spin from the orientation of the longer axis on the inertial reference frame, respectively $3304.356301$ rad/y (a) and $3303.673150$ rad/y (b) for Janus, and $3302.7839016$ (c) and $3305.24602$ (d) for Epimetheus.\label{fig:forlongi}}
\end{figure}

 As explained earlier, we also computed the physical longitudinal librations (Fig.\ref{fig:forlongi}) on the time intervals $[1998.5:2001.7]$ and $[2002.5:2005.7]$, i.e. in excluding the orbital swaps, in considering two different values of the mean orbital motions (Tab.\ref{tab:meanmotion}). The most striking is here the presence of a long-period contribution (probably 8 years), showing the influence of the orbital swap on the rotation. We can also see some short-periodic variations of the amplitudes. Unfortunately, a frequency analysis over such a short time span cannot split contributions alike $\lambda\pm\phi-\varpi$ from $\lambda-\varpi$, whereas it was possible for the tidal librations over a wider interval of study (Tab.\ref{tab:longijan} and \ref{tab:longiepi}).


\par In fact, a frequency analysis would only give a kind of averaged short-period libration (Tab.\ref{tab:forlongi}). The variations of the amplitudes associated can be guessed in reading the plots.


\begin{table}[ht]
\centering
\caption[Longitudinal physical librations]{Longitudinal physical librations $\gamma$. \label{tab:forlongi}}
\begin{tabular}{l|cc|cc}
\tableline\tableline
 & \multicolumn{2}{c}{$[1998.5:2001.7]$} & \multicolumn{2}{c}{$[2002.5:2005.7]$} \\
 & Frequency & Amplitude & Frequency & Amplitude \\
 & rad/y & & rad/y & \\
\tableline
Janus & $3291.2598$ & $0.337\pm0.03^{\circ}$ & $3290.5893$ & $0.338\pm0.03^{\circ}$ \\
Epimetheus & $3289.7159$ & $8.615\pm0.9^{\circ}$ & $3292.1325$ & $8.588\pm0.9^{\circ}$ \\
\tableline
\end{tabular}
\end{table}

\subsection{Analytical validations}

\par Many analytical studies exist on the synchronous rotation. We here propose to use them to validate our numerical results.

\begin{itemize}

\item \emph{Mean obliquity}

\par As explained in the Appendix, the Third Cassini Law implies that the normal to the reference plane (that should be an acceptable Laplace Plane), the normal to the orbital plane and the angular momentum of Janus / Epimetheus are coplanar. As a consequence, 4 equilibria known as Cassini States are possible, in a simplified model. Another analytical formula for the Cassini State 1 (i.e. the most probable one) is given by \citet{hs2004}. In that paper, the equilibrium obliquity $K^*$ related to the orbital plane is given by:

\begin{equation}
\label{equ:Khs}
K^* \approx \frac{\delta_1+\delta_2}{\delta_1+\delta_2-\dot{\ascnode}/n}I
\end{equation}
with 

\begin{equation}
\label{equ:delta1}
\delta_1=-\frac{3}{2}J_2\frac{M\mathcal{R}^2}{C}
\end{equation}
and

\begin{equation}
\label{equ:delta2}
\delta_2=-3C_{22}\frac{M\mathcal{R}^2}{C},
\end{equation}
what gives straightforwardly, using $\epsilon=K^*-I$:

\begin{equation}
\label{equ:oblikhs}
\epsilon=-\frac{I}{1+\frac{3}{2}\frac{n}{\dot{\ascnode}}\frac{J_2+2C_{22}}{C/M\mathcal{R}^2}}.
\end{equation}
This last formula is very similar to the one given in the Appendix for the Cassini State 1 in assuming that $\sin I \approx I$ and $\cos I \approx 1$, which holds for $I<<1$.


\begin{table}[ht]
\centering
\caption[Theoretical Cassini States]{Theoretical Cassini States for Janus and Epimetheus, compared with our numerical simulations. The calculations use the equations given in the Appendix.\label{tab:lescassini}}
\begin{tabular}{l|cc}
\tableline\tableline
 & Janus & Epimetheus \\
\tableline
$\alpha$ (y$^{-1}$) & $1161.73$ & $1542.84$ \\
$\alpha/\dot{\ascnode}$ & $-89.099$ & $-118.315$ \\
$(\alpha/\dot{\ascnode})_{crit}$ & $-1.030$ & $-1.051$ \\
Cassini State 1 & $6.719$ arcsec & $10.834$ arcsec \\
Cassini State 2 & $-89.357^{\circ}$ & $-89.516^{\circ}$ \\
Cassini State 3 & $179.998^{\circ}$ & $179.997^{\circ}$ \\
Cassini State 4 & $89.357^{\circ}$ & $89.516^{\circ}$ \\
Numerical simulations & $5.945$ arcsec & $10.829$ arcsec \\
\tableline
\end{tabular}
\end{table}

\par The table \ref{tab:lescassini} gathers the locations of the theoretical Cassini States and recalls the mean obliquities that we get from our numerical simulations. We can notice that our results are close to the Cassini State 1. We get for Janus significant differences between our values and the analytical ones, while the agreement is very good for Epimetheus. These values have been computed in neglecting the $1:1$ orbital resonance, and in assuming that the orbits of the satellites were circular and uniformly precessing. A possible explanation of the discrepancy observed for Janus could be the influence of a long period that required to fit a slope on the obliquity of Janus (Tab.\ref{tab:obli2}).

\item \emph{Fundamental frequencies of the free librations}

\par We can use previous studies to estimate analytically the three proper frequencies of the librations about the equilibrium. \citet{hs2004} detailed an extensive derivation of the Hamiltonian (\ref{equ:Htotal}) for the satellites in $1:1$ spin-orbit resonance, in assuming that the orbit of the satellite was circular and uniformly precessing, but in considering the three degrees of freedom of the system. After determination of the equilibrium and several canonical transformations, they got the following Hamiltonian:

\begin{equation}
\label{equ:Nquadra}
\mathcal{N}(u,v,w,U,V,W)=\omega_uU+\omega_vV+\omega_wW,
\end{equation}
in which the constants $\omega_u$, $\omega_v$ and $\omega_w$ are the free frequencies of the small librations about the equilibrium. This analytical method has been successfully used in \citep{h2005io,h2005eu,nlv2008}

\par In a more simplified model (i.e. no wobble and a negligible obliquity), we can consider that the libration angle $\gamma$ about the synchronous rotation is ruled by (e.g. \citet{gp1966}):

\begin{equation}
\label{equ:goldreich}
C\ddot{\gamma}+\frac{3}{2}n^2(B-A)H(1,e)\sin 2\gamma=0,
\end{equation}
with $H(1,e)=1-5e^2/2+13e^4/16+O(e^6)$. Since $\gamma$ and the eccentricity $e$ are small, the equation (\ref{equ:goldreich}) becomes:

\begin{equation}
\label{equ:goldreich2}
\ddot{\gamma}+\omega_u^2\gamma=0,
\end{equation}
with

\begin{equation}
\label{equ:omegu}
\omega_u=n\sqrt{3\frac{B-A}{C}}=2n\sqrt{\frac{3C_{22}}{C/M\mathcal{R}^2}}.
\end{equation}
The equation (\ref{equ:goldreich2}) is known to rule a pendulum swinging at the frequency $\omega_u$, i.e. the frequency of the free longitudinal librations. We propose in the Appendix a derivation of the free librations of the obliquity.


\begin{table}[ht]
\centering
\caption[Comparison between the periods of the free librations obtained numerically and in using analytical formulae]{Comparison between the periods (in days) of the free librations obtained numerically and in using analytical formulae. The column "HS 2004" refers to the analytical works of \citet{hs2004}, while "Analytical 1" refers to Eq.(\ref{equ:omegu}) for $T_u=2\pi/\omega_u$ and to Eq.(\ref{equ:tv}) for $T_v$. "Analytical 2" refers to to Eq.(\ref{equ:tv2}) for $T_v$.\label{tab:companafrek}}
\begin{tabular}{c|cccc}
\tableline\tableline
 & Numerical & HS 2004 & Analytical 1 & Analytical 2 \\
\tableline
Janus & & & & \\
\tableline
$T_u$ & $1.26713$ & $1.26053$ & $1.26814$ & - \\
$T_v$ & $2.17884$ & $1.98573$ & $1.99787$ & $1.76767$ \\
$T_w$ & $2.09798$ & $1.74391$ & - & - \\
\tableline
Epimetheus & & & & \\
\tableline
$T_u$ & $0.74717$ & $0.73255$ & $0.73697$ & - \\
$T_v$ & $1.80386$ & $1.49107$ & $1.50015$ & $1.49946$ \\
$T_w$ & $5.54234$ & $4.71920$ & - & - \\
\tableline
\end{tabular}
\end{table}

\par The Tab.\ref{tab:companafrek} gives a comparison between our numerical values of the periods of the free librations and values due to formulae obtained by analytical studies.  The column "Analytical 2" uses the Eq.(\ref{equ:tv2}) that uses the value of obliquity coming from our simulations. We can see a good agreement for the period of the free librations in longitude $T_u$, but a significant discrepancy for the two other periods, that the use of the real obliquity does not reduce.  This discrepancy is probably due to the approximation on the dynamics of the satellites, in particular the effects of the $1:1$ orbital resonance.

\item \emph{Longitudinal librations}

\par We have defined two kinds of longitudinal librations: the physical ones $\gamma$, and the tidal ones $\psi$. We have (e.g. \citet{md1999}):

\begin{equation}
\label{equ:libfor}
\gamma=\frac{2e}{1-(n/\omega_u)^2}\sin(nt+\varphi),
\end{equation}
where $\varphi$ is a constant phase depending on the time origin. This formula assumes that the orbit is close to be keplerian and circular, in particular its eccentricity is assumed to be small and constant. The tidal librations $\psi$, representing the misalignment of the tidal bulge of the satellite, is obtained in considering the \emph{optical libration} $\phi=2e\sin nt=\gamma-\psi$, due to the variations of the velocity of the satellite on its orbit. We then find \citep{ttb2009}:

\begin{equation}
\label{equ:libtid}
\psi=\frac{-2e}{1-(\omega_u/n)^2}\sin(nt+\varphi).
\end{equation}


\begin{table}[ht]
\centering
\caption[Numerical and analytical determinations of the librations]{Comparison between the numerical and analytical determinations of the librations of Janus and Epimetheus. The column "Numerical" gathers the results given by the Tab.\ref{tab:err} and \ref{tab:forlongi}, while the "Analytical" ones use the formulae (\ref{equ:libfor}) and (\ref{equ:libtid}) with two different values of the proper frequency $\omega_u$: the numerical one in column "1", and the one given by the formula (\ref{equ:omegu}) in the last column. These calculations have been made using the eccentricities at J2000.0 that we determined in fitting a slope, i.e. $7.30539\times10^{-3}$ for Janus and $1.01621\times10^{-2}$ for Epimetheus.\label{tab:compamplibra}}
\begin{tabular}{c|rrr}
\tableline\tableline
 & Numerical & Analytical 1 & Analytical 2 \\
\tableline
Janus & & & \\
\tableline
$\gamma$ & $\approx-0.34\pm0.03^{\circ}$  & $-0.35959^{\circ}$  & $-0.35877^{\circ}$  \\
$\psi$ & $\approx-1.1^{\circ}$ & $-1.19673^{\circ}$ & $-1.19591^{\circ}$ \\
\tableline
Epimetheus & & & \\
\tableline
$\gamma$ & $-8.6\pm0.9^{\circ}$ & $-7.41096^{\circ}$ & $-9.26011^{\circ}$ \\
$\psi$ & $\approx-10^{\circ}$ & $-8.57545^{\circ}$ & $-10.42460^{\circ}$ \\
\tableline
\end{tabular}
\end{table}

\par The Tab.\ref{tab:compamplibra} gives a comparison between the amplitudes of the librations that we detected numerically, and the ones given by the analytical formulae (\ref{equ:libfor}) and (\ref{equ:libtid}), in using the proper period $T_u$ detected numerically (column "1") and analytically (column "2"). For Janus, we can see a quite good agreement between the numerical and analytical values, even if the discrepancy is significant. However, an agreement is not so obvious for Epimetheus, even between the analytical values. The reason is probably a high sensitivity on the gravity field parameters ($J_2$, $C_{22}$ and $C/M\mathcal{R}^2$) probably because, as already indicated by \citet{ttb2009}, Epimetheus seems to be close to the $1:1$ secondary resonance between its spin and its free longitudinal librations.

\end{itemize}

\section{Discussion}

\par Recently, \citet{ttb2009} observed the rotation of Janus and Epimetheus thanks to the Cassini spacecraft. We here propose to use the theory of the rotation we have elaborated to try to explain the observations, and to get planetological consequences that could be observed.

\subsection{Uncertainty of the solutions}

\par All our calculations assume that the input parameters are known with a very high accuracy. In fact, \citet{ttb2009} derived them from fits to observations, giving most probable values and uncertainties (see Tab.\ref{tab:observ}). A rigorous way to study the influence of the uncertainties of the input parameters on the rotation would be to run the numerical integrations with some variations of these parameters within the error bars. This method requires lots of CPUs and would give uncertainties on useless values, like amplitudes of some periodic contributions that are far too small to be observed. That is the reason why we propose to use the analytical formulae to derive uncertainties on the longitudinal librations and the equilibrium obliquities.


\begin{table}
\centering
\caption{Parameters derived from Cassini observations \citep{ttb2009}. \label{tab:observ}}
\begin{tabular}{l|cccc}
\tableline\tableline
 & $(B-A)/C$ & $a$ (km) & $b$ (km) & $c$ (km) \\
\tableline
Janus & $0.100\pm0.012$ & $101.5\pm1.9$ & $92.5\pm1.2$ & $76.3\pm1.2$ \\
Epimetheus & $0.296^{+0.019}_{-0.027}$ & $64.9\pm2.0$ & $57.0\pm3.7$ & $53.1\pm0.7$ \\
\tableline
\end{tabular}
\end{table}

\par For this, two parameters need to be known for each body: $(B-A)/C$ for the amplitude of the longitudinal librations (Eq.\ref{equ:libfor} and \ref{equ:libtid}), and $(C-A)/C$ (Eq.\ref{equ:alfa}) for the equilibrium obliquity. The uncertainties on $(B-A)/C$ are given by \citet{ttb2009} and so can be used directly, while the ones on $(C-A)/C$ should be guessed from the uncertainties on the dimensions of the bodies. In assuming that they are homogeneous ellipsoids, we have

\begin{equation}
\label{equ:moments}
A\approx\frac{M}{5}(b^2+c^2),B\approx\frac{M}{5}(a^2+c^2),C=\frac{M}{5}(a^2+b^2),
\end{equation}
what yields

\begin{equation}
\label{equ:cma}
\frac{C-A}{C}\approx\frac{a^2-c^2}{a^2+b^2}.
\end{equation}

\par Unfortunately, this last formula does not work for Epimetheus because, as \citet{ttb2009} noticed, it in fact deviates significantly from an ellipsoid. Moreover, the moments of inertia given by \citet{ttb2009} are taken directly from a numerical shape model derived from the data \citep{t1993,stcv1993,tdcosncmsb1998}, and are not directly related to the best-fit ellipsoid. We also had troubles in using it, getting obliquities far bigger than the mean obliquity ($10.8$ arcsec) we give in Tab.\ref{tab:obli2} or the maximum obliquity $14$ arcsec (Tab.\ref{tab:err}). So, we used it only for Janus and got $(C-A)/C=0.234^{+0.032}_{-0.025}$. The Tab.\ref{tab:uncert} gathers our results.


\begin{table}[ht]
\centering
\caption[Uncertainty on the output variables]{Uncertainty on the longitudinal librations and on the obliquity, deduced from the uncertainty on the parameters given by \citet{ttb2009}. \label{tab:uncert}}
\begin{tabular}{l|cc}
\tableline\tableline
 & Janus & Epimetheus \\
\tableline
$\gamma$ $(^{\circ})$ & $-0.36\pm0.06$ & $-9.2^{+4.3}_{-10.8}$ \\
$\psi$ $(^{\circ})$ & $-1.2\pm0.06$ & $-10.4^{+4.4}_{-10.8}$ \\
Mean obliquity (arcsec) & $6.72\pm0.82$ & -- \\
\tableline
\end{tabular}
\end{table}

\par
\begin{figure}[ht]
\plotone{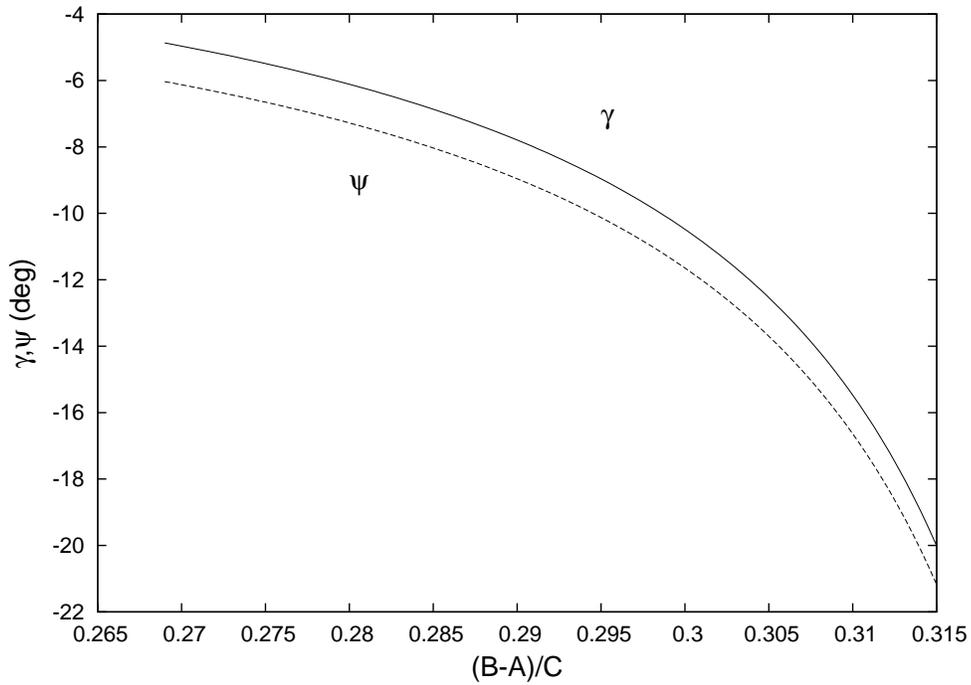}
\caption[Sensitivity of the librations of Epimetheus]{Dependency of the longitudinal librations of Epimetheus on the moments of inertia. We can notice a high sensitivity due to the proximity of a resonance.\label{fig:errepime}}
\end{figure}

 \citet{ttb2009} predicted the physical librations $\gamma$ as $-0.33\pm0.06^{\circ}$ for Janus and $-8.9^{+4.2}_{-10.4}$ $^{\circ}$ for Epimetheus, from shape-derived moments of inertia that assume constant density. We have good agreements, the differences being likely to be due to different considered values for the eccentricities. As already said, the longitudinal motion of Epimetheus is very sensitive to the input parameters because of the proximity of the $1:1$ secondary resonance between the spin and the free longitudinal librations. This way, when the input parameter $(B-A)/C$ gets closer to the critical value $1/3$, the amplitude of the libration increases dramatically (Fig.\ref{fig:errepime}). We have a similar phenomenon for the longitudinal librations of Mercury, close to a resonance with the mean motion of Jupiter \citep{pym2009}.

\placefigure{fig:errepime}

\subsection{Comparison with the observations}

\par The main results of \citet{ttb2009} on the rotation of Janus and Epimetheus are detections of the physical librations of Janus and Epimetheus (cf. Tab.\ref{tab:compobserv}), and a significant offset of Janus' sub-Saturn point from the axis of the minimum moment of inertia.


\begin{table}[ht]
\centering
\caption[Comparison between the observations and our results]{Comparison between the observations of \citet{ttb2009} and our results. The column "our simulations" refers to the results of the numerical integration (Tab.\ref{tab:forlongi} and \ref{tab:obli2}) while mean values refers to the study of the uncertainties (Tab.\ref{tab:uncert}).\label{tab:compobserv}}
\begin{tabular}{l|ccc}
\tableline\tableline
 & \citet{ttb2009} & Our simulations & Mean values \\
\tableline
Janus & & & \\
\tableline
$\gamma$ & $-0.3\pm0.9^{\circ}$ & $-0.34\pm0.03^{\circ}$ & $-0.36\pm0.06^{\circ}$ \\
Obliquity $\epsilon$ & -- & $5.95$ arcsec & $6.72\pm0.82$ arcsec \\
Longitudinal offset & $5.2\pm1^{\circ}$ & -- & $0$ \\
Latitudinal offset & $2.3\pm1^{\circ}$ & -- & $0$ \\
\tableline
Epimetheus & & & \\
\tableline
$\gamma$ & $-5.9\pm1.2^{\circ}$ & $-8.6\pm0.9^{\circ}$ & $-9.9^{+4.9}_{-13.4}$ $^{\circ}$ \\
Obliquity $\epsilon$ & -- & $10.83$ arcsec & -- \\
Longitudinal offset & $<1^{\circ}$ & -- & $0$ \\
Latitudinal offset & $<1^{\circ}$ & -- & $0$ \\
\tableline
\end{tabular}
\end{table}

\par Our study seems to agree with the observed value of the physical longitudinal librations $\gamma$ for Janus, even if we should keep in mind that the error bars of the observations are thrice bigger than the most probable value, so we cannot be certain that this libration has actually been observed. For Epimetheus, there is a significant discrepancy on the mean values of $\gamma$, but with an overlap due to the sensitivity on the input parameters. An interesting point is the error bar coming from our numerical simulations, that assumed the input parameters to be exactly known. This error bar is due to the variations of the amplitude that we read on the plots (Fig.\ref{fig:forlongi}). The variation of $0.9^{\circ}$ could partly explain the error bar of $1.2^{\circ}$ observed by \citet{ttb2009}. The expected obliquities are probably too small to be detected.

\par Our model does not explain the offset of Janus' sub-Saturn point from the axis of the moment of inertia. If we suppose it could be due to the orbital swap or to the latitudinal librations, it should anyway be lower than $40$ arcmin in longitude (Fig.\ref{fig:forlongi}) and that $6.5$ arcsec in latitude (Tab.\ref{tab:err}). One way to get it might be to introduce non-diagonal terms in the matrix of inertia (Eq.\ref{equ:inertie}). This would physically mean that the geometrical principal axes differ significantly from the gravitational ones, so that Janus is not an homogenous body, or that its equipotential surface differs significantly from an ellipsoid. We leave this for future work.

\section{Conclusion}

\par We have elaborated in this paper a 3-degree of freedom theory of the rotation of Janus and Epimetheus, based on the physical parameters given by \citet{ttb2009}, derived from Cassini observations. This study used an accurate representation of the orbital motions of these bodies, so as to evaluate the rotational consequences of the orbital swaps.

\par Our numerical integrations show that these swaps induce significant variations of the short-period longitudinal librations, and thus could explain the error bars on the observations, at least for Epimetheus. On the contrary, this model does not explain the mean orientation of Janus' axis of minimum moment of inertia. Moreover, we give an estimation of some undetected aspects of the rotation, like the latitudinal librations ($\approx6.5$ arcsec for Janus and $\approx11$ arcsec for Epimetheus) and the mean obliquities (respectively $6$ and $11$ arcsec). Finally we highlight, as already \citet{ttb2009} did, the sensitivity of the longitudinal motion of Epimetheus on the input parameters, because of the proximity of a resonance.

\par It will be probably possible, in a next future, to get a better accuracy on the observations, particularly thanks to next fly-bys. We hope that it will open the door to an inversion of the theory of the rotation to get several parameters of the gravity field of these bodies, and so information on their interior.

\section*{Acknowledgments}

\par The author is indebted to Laetitia Legrain, Dimitri Tomanos and Vincent Malmedy for their help on the interpolation of the JPL HORIZONS tables, and Sandrine D'Hoedt, Nicolas Delsate, Julien Dufey and Philippe Robutel for fruitful discussions.

\appendix

\section{Approximation of the mean Laplace Plane}

\par The Lagrange equation ruling the precessional motion of Janus/Epimetheus is, at the first approximation (i.e. in neglecting the effects of eccentricities and inclinations)

\begin{equation}
\label{equ:ascjanus}
\frac{d\ascnode}{dt}=\omega_{\saturn}+\sum_i\omega_i+\omega_{\odot}
\end{equation}
with 

\begin{equation}
\label{equ:omegsat}
\omega_{\saturn}=-\frac{3}{2}J_2n\Big(\frac{R_{\saturn}}{a}\Big)^2,
\end{equation}

\begin{equation}
\label{equ:omegi}
\omega_{i}=-\frac{1}{4}\frac{M_i}{M_{\saturn}}\bigg(\frac{a}{a_i}\bigg)^2b_{3/2}^{(1)}\bigg(\frac{a}{a_i}\bigg),
\end{equation}
and

\begin{equation}
\label{equ:omegsun}
\omega_{\odot}=-\frac{1}{4}\frac{M_{\odot}}{M_{\saturn}}\bigg(\frac{a}{a_{\odot}}\bigg)^2b_{3/2}^{(1)}\bigg(\frac{a}{a_{\odot}}\bigg).
\end{equation}
(see e.g. \citet{cv1999}), where $n$ is Janus/Epimetheus' mean motion, $a_i$ the semimajor axis of satellite $i$ (1 to 8 standing respectively for Mimas to Iapetus, the other ones being neglected), $M_i$ its mass, $a_{\odot}$ and $M_{\odot}$ the same quantities for the Sun, $\mathcal{R}_{\saturn}$ the equatorial radius of Saturn, and $b_{3/2}^{(1)}(x)$ is a classical Laplace coefficient. The Laplace coefficients are computed with the formula (see e.g. \citet{bc1960}):

\begin{equation}
\label{equ:laplacoeff}
b_{3/2}^{(1)}(x)=\frac{2}{\pi}\int_0^{\pi}\frac{\cos\theta}{(1-2x\cos\theta+x^2)^{3/2}}d\theta.
\end{equation}

\par We have \citep{ym2006}

\begin{equation}
\label{equ:p0}
p_0=\frac{\sum_ip_i\omega_i+p_{\odot}\omega_{\odot}+p_{\saturn}\omega_{\saturn}}{\sum_i\omega_i+\omega_{\odot}+\omega_{\saturn}}
\end{equation}
and

\begin{equation}
\label{equ:q0}
q_0=\frac{\sum_iq_i\omega_i+q_{\odot}\omega_{\odot}+q_{\saturn}\omega_{\saturn}}{\sum_i\omega_i+\omega_{\odot}+\omega_{\saturn}},
\end{equation}
where $p_k=\sin\ascnode_k\sin i_k$ and $q_k=\cos\ascnode_k\sin i_k$ for the satellites 1 to 8, the Sun and Saturn, $i_k$ standing for the inclination of the satellite $k$, and $\ascnode_k$ for its ascending node. $p_0$ and $q_0$ indicate the location of the Laplace Plane. The numerical values of $\omega$ are gathered in Tab.\ref{tab:lapla}.


\begin{table}[ht]
\centering
\caption[Determination of the Laplace Plane]{Determination of the Laplace Plane of Janus and Epimetheus. The gravity field data and mean semi-major axes come from JPL HORIZONS at J2000, except $\mathcal{G}M_{\odot}$ that come from the recommendations of IERS. The $\omega$ have been computed thanks to Eq.\ref{equ:omegi} and \ref{equ:omegsun}. The semimajor axes are given with Saturn as the central body. The inclinations with respect to Saturn's equatorial plane are estimated from the main terms of the series of TASS1.7 \citep{vd1995,dv1997}, this rough estimation being here accurate enough because we just want to show that the inclination of the Laplace Plane is small. These values are easier to use than HORIZONS' because HORIZONS give them with respect of the Laplace Plane of the considered body. \label{tab:lapla}}
\begin{tabular}{l|cccc}
\tableline\tableline
 & $\mathcal{G}M$ & $a$ & $\omega$ & $\sin i_k$\\
 & $km^3.s^{-2}$ & $km$ & & \\
\tableline
Mimas & $2.5026$ & $185539$ & $-9.15\times10^{-7}$ & $2.4\times10^{-2}$\\
Enceladus & $7.2027$ & $238037$ & $-4.01\times10^{-7}$ & $2.5\times10^{-4}$ \\
Tethys & $41.2067$ & $294672$ & $-8.03\times10^{-7}$ & $1.6\times10^{-2}$ \\
Dione & $73.1146$ & $377415$ & $-5.27\times10^{-7}$ & $3.3\times10^{-4}$ \\
Rhea & $153.9426$ & $527068$ & $-3.44\times10^{-7}$ & $5.9\times10^{-3}$ \\
Titan & $8978.1382$ & $1221865$ & $-1.41\times10^{-6}$ & $1.1\times10^{-2}$ \\
Hyperion & $0.3727$ & $1500934$ & $-3.12\times10^{-11}$ & $1.2\times10^{-2}$ \\
Iapetus & $120.5038$ & $3560851$ & $-7.44\times10^{-10}$ & $0.26$ \\
Sun & $1.32712442076\times10^{11}$ & $1.4266641409\times10^9$ & $-1.27\times10^{-8}$ & $0.47$ \\
Saturn & $3.7931208\times10^7$ & $0$ & $-12.71$ & $0$ \\
\tableline
\end{tabular}
\end{table}

\par A rigorous determination of the instantaneous Laplace Plane would require to consider the time variations of the quantities $p$ and $q$ of each perturber, to get $p_0(t)$ and $q_0(t)$, that could be finally averaged to get a mean Laplace Plane, that would be an optimized inertial reference plane. Since we just want to show that the mean Laplace Place is very close to the Saturnian equator at J2000, we will give an upper bound for the inclination of the Laplace Plane $i_0$, in considering that

\begin{equation}
\sin i_0<\frac{\sum_i\sin i_i\omega_i+\sin i_{\odot}\omega_{\odot}+\sin i_{\saturn}\omega_{\saturn}}{\sum_i\omega_i+\omega_{\odot}+\omega_{\saturn}},
\label{equ:izero}
\end{equation}
and we get numerically $i_0<4.63\times10^{-9}$ rad, i.e. $\approx 1$ milli-arcsec.

\section{The Cassini states}

\par \citet{c1966} showed that the spin axis of a rotating body has 2 or 4 equilibria named Cassini states. We give here an analytical study of the location of these states, widely inspired from \citep{wh2004}, but rewritten our way. The reader can find alternative explanations in the literature (e.g. \citet{p1969,b1972,hm1987,dlr2006}).

\par We here consider a reference plane of normal $\vec{k}$, and we assume that the orbital plane has a constant inclination $I$ and precesses at a uniform rate 
$\dot{\ascnode}$. Under these assumptions, the reference plane can be considered as a Laplace Plane. We name $\vec{n}$ the normal to the orbit and $\vec{s}$ the unit vector colinear to the angular momentum of the considered body. Finally, we call $\epsilon$ the obliquity of the body, it is the angle between $\vec{s}$ and $\vec{n}$. The Third Cassini Law tells us that the vectors $\vec{k}$, $\vec{n}$ and $\vec{s}$ are coplanar, so we say that the angle between $\vec{k}$ and $\vec{s}$ is $I+\epsilon$. This convention on the orientation of $\epsilon$ is consistent with our model, \citet{wh2004} chose another one.

\par The equation of motion ruling $\vec{s}$ reads:

\begin{equation}
\label{equ:ward}
\frac{d\vec{s}}{dt}=\alpha(\vec{s}\cdot\vec{n})(\vec{s}\times\vec{n})+\dot{\ascnode}(\vec{s}\times\vec{k}),
\end{equation}
where $\alpha$ is a precessional constant that can be written \citep{w1975}:

\begin{equation}
\label{equ:alfa}
\alpha=\frac{3}{2}\frac{(C-A)n^2}{C\omega}=\frac{3}{2}\frac{n^2}{\omega}\frac{J_2+2C_{22}}{C/MR^2},
\end{equation}
where $n$ is the orbital mean motion and $\omega$ is the spin velocity of the body. For most of the natural satellites like Janus and Epimetheus, the synchronous rotation induces $n=\omega$, and we get

\begin{equation}
\label{equ:alfa2}
\alpha=\frac{3}{2}n\frac{J_2+2C_{22}}{C/MR^2}.
\end{equation}

\par At the equilibrium, we have $\frac{d\vec{s}}{dt}=0$ in the reference frame defined by the vector $\vec{n}$, $\vec{k}$ and their product. In fact, $\vec{n}$ and $\vec{s}$ precess synchronously. The projection of the equation (\ref{equ:ward}) on the direction normal to the plane $(\vec{s},\vec{k})$ yields:

\begin{equation}
\label{equ:csavant}
\frac{\alpha}{2\dot{\ascnode}}\sin 2\epsilon+\sin(I+\epsilon)=0.
\end{equation}

\par It is well known that, for most of the natural satellites and Mercury, the slow precessional motion induces the existence of four equilibria that are close respectively to $0$, $\frac{\pi}{2}$, $\pi$, $-\frac{\pi}{2}$. These equilibria are known as Cassini States 1, 4, 3 and 2. With a more rapid precessional motion, three of these equilibria get closer (except Cassini State 3), and two of them can vanish as it is the case for the Moon. 

\begin{figure}[ht]
\centering
\plottwo{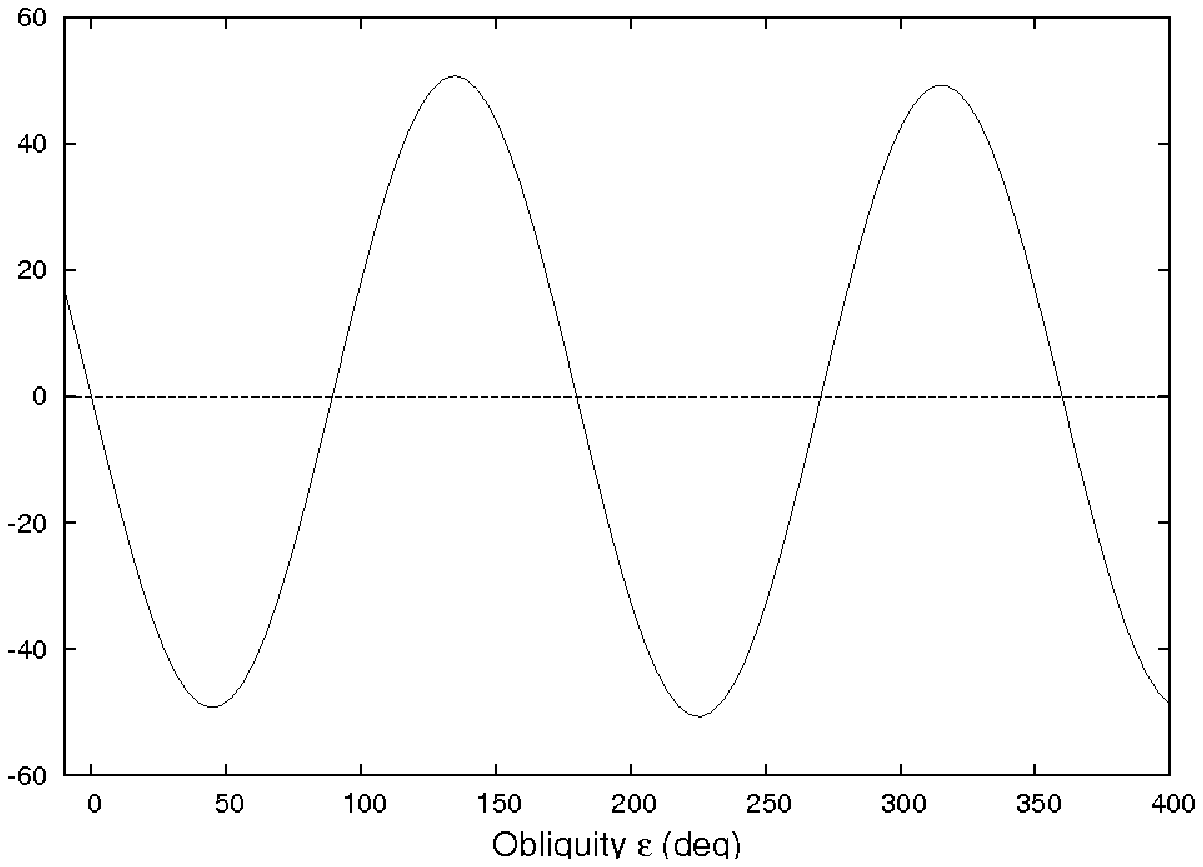}{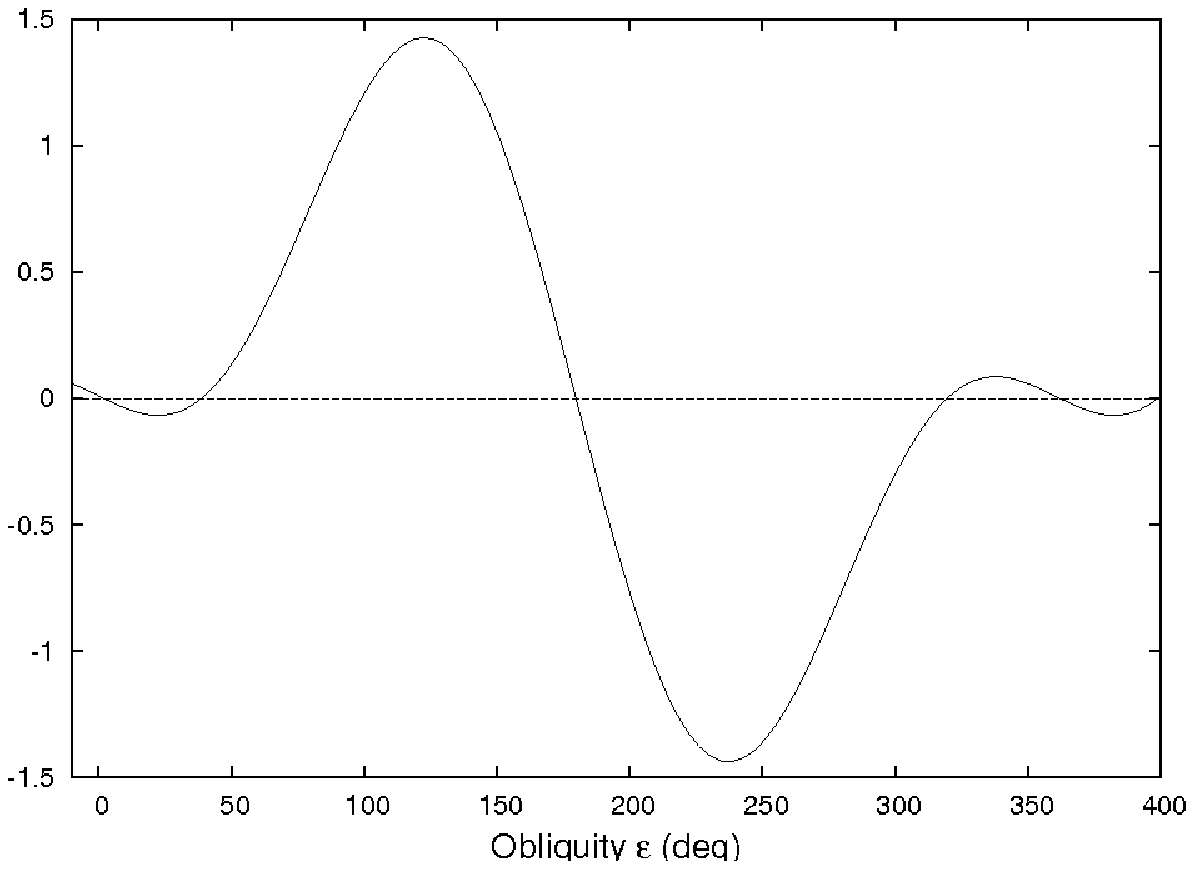}
\caption[Locations of the Cassini States]{Locations of the Cassini States from the equation (\ref{equ:csavant}), respectively with $\alpha/\dot{\ascnode}=-100$ (left), and $\alpha/\dot{\ascnode}=-1.3$ (right), for an orbital inclination $I=0.01$ rad. The y axis is the quantity $\alpha/(2\dot{\ascnode})\sin 2\epsilon+\sin(I+\epsilon)$, the location of the Cassini States being its roots. The 4 Cassini States are respectively, from left to right : 1, 4, 3 and 2. We can see that for a high value of the ratio $|\alpha/\dot{\ascnode}|$ the four equilibria are close to multiples of $\pi/2$, while the states 1, 4 and 2 get closer when this ratio tends to 1.\label{fig:cs}}
\end{figure}

\placefigure{fig:cs}

\par As can be seen on Fig.\ref{fig:cs}, the value of $\alpha/\dot{\ascnode}$ is critical. It can be shown \citep{wh2004} that the 4 Cassini States exist if $|\alpha/\dot{\ascnode}|>(\sin^{2/3}I+\cos^{2/3}I)^{3/2} \approx 1$. For the "classical" case of the natural satellites where $|\alpha/\dot{\ascnode}|\gg 1$, it is straightforward to develop the equation (\ref{equ:csavant}) around $0$, $\frac{\pi}{2}$, $\pi$, $-\frac{\pi}{2}$ and we get the following formulae:

\begin{displaymath}
\begin{array}{ll}
  \epsilon  \approx -\frac{\sin I}{\alpha/\dot{\ascnode}+\cos I},  & \textrm{(Cassini State 1)} \\
  \epsilon \approx -\frac{\pi}{2}+\frac{\cos I}{\sin I-\alpha/\dot{\ascnode}},  & \textrm{(Cassini State 2)} \\
  \epsilon \approx \pi+\frac{\sin I}{\alpha/\dot{\ascnode}-\cos I}, &  \textrm{(Cassini State 3)} \\
  \epsilon \approx \frac{\pi}{2}+\frac{\cos I}{\sin I+\alpha/\dot{\ascnode}}.  & \textrm{(Cassini State 4)}
\end{array}
\end{displaymath}

\section{Free librations of the obliquity}

\par The aim of this section is to estimate the period of the small free libration of the obliquity about the exact Cassini State 1. We start from the following Hamiltonian \citep{w1975,wh2004}, representing the behavior of the obliquity, that is in fact a part of the Hamiltonian of the system:

\begin{equation}
\label{equ:hamilk}
\mathcal{H}=-\frac{\alpha}{2}(\vec{s}\cdot\vec{n})^2-\dot{\ascnode}(\vec{s}\cdot\vec{k}),
\end{equation}
where $\alpha$, $\vec{s}$, $\vec{n}$ and $\vec{k}$ are defined as above. We then find:

\begin{equation}
\label{equ:hamilk2}
\mathcal{H}=-\frac{\alpha}{2}\cos^2\epsilon-\dot{\ascnode}\cos(I+\epsilon)
\end{equation}
and, at the second order in $\epsilon$:

\begin{equation}
\label{equ:hamilk3}
\mathcal{H}\approx\epsilon\dot{\ascnode}\sin I+\frac{\epsilon^2}{2}(\alpha+\dot{\ascnode}\cos I).
\end{equation}

\par In order to study the small variations of the obliquity $\epsilon$, we set $\epsilon=\epsilon_0+\epsilon_1$, where $\epsilon_0$ is the constant obliquity at the equilibrium, and $|\epsilon_1|\ll|\epsilon_0|$. We can now write, in dropping the constant terms:

\begin{eqnarray}
\mathcal{H} & = & \epsilon_1\big(\epsilon_0(\alpha+\dot{\ascnode}\cos I)+\dot{\ascnode}\sin I\big)+\frac{\epsilon_1^2}{2}(\alpha+\dot{\ascnode}\cos I) \nonumber \\
 & = & \frac{\epsilon_1^2}{2}(\alpha+\dot{\ascnode}\cos I) \label{equ:hamilk4}
\end{eqnarray}
because the definition of the Cassini State 1 gives $\epsilon_0(\alpha+\dot{\ascnode}\cos I)+\dot{\ascnode}\sin I=0$. We have now a quadratic function of $\epsilon_1$.

\par $\epsilon_1$ is not really a canonical variable. It results from the classical polar transformation of canonical variables that the Hamiltonian (\ref{equ:hamilk4}) can be written as

\begin{equation}
\label{equ:hamilk5}
\mathcal{H}(v,V)=\omega_v V
\end{equation}
where $v$ and $V$ are canonical angle-action variables, $V=\frac{\epsilon_1^2}{2}$ and $\omega_v=\alpha+\dot{\ascnode}\cos I$ is the frequency of the free oscillations. So, the period of the free librations of the obliquity is, in this simplified model:

\begin{equation}
\label{equ:tv}
T_v=\frac{2\pi}{\alpha+\dot{\ascnode}\cos I},
\end{equation}
and can also be written as

\begin{equation}
\label{equ:tv2}
T_v=-\frac{2\pi\epsilon_0}{\dot{\ascnode}\sin I}.
\end{equation}

\section{Supplemental Material}

\subsection{Orbital motion}

Here are the frequency analyses of the orbital elements of Janus and Epimetheus.







\begin{table}[ht]
 \centering
\caption{Mean longitude of Janus, after removal of the proper mode $\lambda$. The series are in sine.\label{tab:lambjan}}
\begin{tabular}{r|rrrrr}
\tableline\tableline
 N & $\phi$ & $\omega$ & Amplitude (rad) & Period \\
\tableline
 $1$ & $1$  & -   &  $0.55346$ &   $8.00687$ y \\
 $2$ & $3$  & -   & $-0.06351$ &   $2.66892$ y \\
 $3$ & $5$  & -   &  $0.02017$ &   $1.60135$ y \\
 $4$ & -    & $2$ &  $0.00798$ &  $68.12$ y \\
 \tableline
 \end{tabular}
\end{table}

\begin{table}[ht]
 \centering
\caption{$z=k+\sqrt{-1}h=e\exp(\sqrt{-1}\varpi)$ for Janus. The series are in complex exponentials.\label{tab:zjan}}
\begin{tabular}{r|rrrrr}
\tableline\tableline
 N & $\lambda$ & $\phi$ & $\varpi_J$ & Amplitude ($\times10^3$) & Period \\
\tableline
 $1$ & -   & -    & $1$ & $6.64720$ & $175.30788$ d \\
 $2$ & $1$ & -    & -   & $3.59821$ &   $0.69459$ d \\
 $3$ & $1$ & $1$  & -   & $1.03092$ &   $0.69442$ d \\
 $4$ & $1$ & $-1$ & -   & $1.03032$ &   $0.69475$ d \\
 $5$ & $1$ & $2$  & -   & $0.17933$ &   $0.69426$ d \\
 $6$ & $1$ & $-2$ & -   & $0.17920$ &   $0.69492$ d \\
 $7$ & $1$ & $3$  & -   & $0.09885$ &   $0.69409$ d \\
 \tableline
 \end{tabular}
\end{table}

\begin{table}[ht]
 \centering
\caption{$\zeta=q+\sqrt{-1}p=\sin\Big(\frac{I}{2}\Big)\exp(\sqrt{-1}\ascnode)$ for Janus. The series are in complex exponentials.\label{tab:zetajan}}
\begin{tabular}{r|rrrr}
\tableline\tableline
 N & $\ascnode_J$ & $\omega$ & Amplitude ($\times10^3$) & Period \\
\tableline
 $1$ & $1$ &  -  & $1.43593$ & $-176.00968$ d \\
 $2$ &   - & $1$ & $0.00797$ &  $136.16498$ y \\
\tableline
  \end{tabular}
\end{table}

\clearpage

\begin{table}[ht]
 \centering
\caption{Mean longitude of Epimetheus, after removal of the proper mode $\lambda-\pi$. The series are in sine.\label{tab:lambepim}}
\begin{tabular}{r|rrr}
\tableline\tableline
 N & $\phi$ & Amplitude (rad) & Period \\
\tableline
 $1$ & $1$ & $-1.99423$ &   $8.00676$ y \\
 $2$ & $3$ &  $0.22886$ &   $2.66892$ y \\
 $3$ & $5$ & $-0.07271$ &   $1.60136$ y \\
 $4$ & $7$ &  $0.03345$ &   $1.14382$ y \\
 $5$ & $9$ & $-0.01828$ & $324.94152$ d \\
\tableline
 \end{tabular}
\end{table}

\begin{table}[ht]
 \centering
\caption{$z=k+\sqrt{-1}h=e\exp(\sqrt{-1}\varpi)$ for Epimetheus. The series are in complex exponentials.\label{tab:zepim}}
\begin{tabular}{r|rrrrr}
 \tableline\tableline
 N & $\lambda$ & $\phi$ & $\varpi_E$ & Amplitude ($\times10^3$) & Period \\
\tableline
 $1$ & -   & -    & $1$ &  $9.82435$ & $175.28140$ d \\
 $2$ & $1$ & $1$  & -   & $-2.07104$ &   $0.69442$ d \\
 $3$ & $1$ & $-1$ & -   & $-2.07007$ &   $0.69475$ d \\
 $4$ & $1$ & $2$  & -   & $-1.59186$ &   $0.69426$ d \\
 $5$ & $1$ & $-2$ & -   & $-1.59002$ &   $0.69492$ d \\
 $6$ & $1$ & -    & -   & $-0.99815$ &   $0.69459$ d \\
 $7$ & $1$ & $3$  & -   & $-0.45071$ &   $0.69409$ d \\
 $8$ & $1$ & $-3$ & -   & $-0.44971$ &   $0.69508$ d \\
 $9$ & $1$ & $4$  & -   & $-0.18343$ &   $0.69393$ d \\
\tableline
 \end{tabular}
\end{table}

\begin{table}[ht]
 \centering
\caption{$\zeta=q+\sqrt{-1}p=\sin\Big(\frac{I}{2}\Big)\exp(\sqrt{-1}\ascnode)$ for Epimetheus. The series are in complex exponentials.\label{tab:zetaepim}}
\begin{tabular}{r|rrrrrr}
\tableline\tableline
 N & $\phi$ & $\ascnode_E$ & $\omega$ & Amplitude ($\times10^3$) & Period \\
\tableline
 $1$ & -    & $1$ & -   & $3.07916$ & $-175.99124$ d \\
 $2$ & $-1$ & $1$ & -   & $0.02859$ & $-166.00121$ d \\
 $3$ &  $1$ & $1$ & -   & $0.02841$ & $-187.26047$ d \\
 $4$ & -    & -   & $1$ & $0.00797$ &  $136.15656$ y \\
\tableline
  \end{tabular}
\end{table}

\subsection{Motion of the North Pole}



\begin{table}[ht]
 \centering
\caption{Motion of the North Pole of Janus' rotation axis about the North Pole of Janus, in meters. The variable represented here is $Q_1+\sqrt{-1}Q_2=\sin J\big(\sin l \big(1+(J_2+2C_{22})/C\big)+\sqrt{-1}\cos l \big(1+(J_2-2C_{22})/C\big)\big)$ (see Eq.\ref{equ:Q1} and \ref{equ:Q2}). The series are in complex exponentials, and the cut-off is at 10 cm.\label{tab:wobijan}}
\begin{tabular}{r|rrrrrr}
\tableline\tableline
 N & $\lambda$ & $\phi$ & $\ascnode_J$ & Amplitude & Phase & Period \\
\tableline
 $1$ &  $1$ & -    & $-1$ & $1.16$ & $-138.717^{\circ}$ &   $0.69186$ d \\
 $2$ & $-1$ & -    &  $1$ & $0.79$ &  $-41.283^{\circ}$ &  $-0.69186$ d \\
 $3$ &  $1$ &  $1$ & $-1$ & $0.32$ & $-140.972^{\circ}$ &   $0.69170$ d \\
 $4$ &  $1$ & $-1$ & $-1$ & $0.32$ &   $43.539^{\circ}$ &   $0.69202$ d \\
 $5$ & $-1$ & $-1$ &  $1$ & $0.23$ &  $-39.029^{\circ}$ &  $-0.69170$ d \\
 $6$ & $-1$ &  $1$ &  $1$ & $0.23$ &  $136.461^{\circ}$ &  $-0.69202$ d \\
 \tableline
\end{tabular}
\end{table}

\begin{table}[ht]
 \centering
\caption{Motion of the North Pole of Epimetheus' rotation axis about the North Pole of Epimetheus, in meters. The series are in complex exponentials, and the cut-off is at 20 cm.\label{tab:wobiepi}}
\begin{tabular}{r|rrrrrrr}
\tableline\tableline
 N & $\lambda$ & $\phi$ & $\varpi_E$ & $\ascnode_E$ & Amplitude & Phase & Period \\
\tableline
 $1$ &  $1$ &  $1$ & -    & $-1$ & $0.72$ & $-179.666^{\circ}$ &   $0.69170$ d \\
 $2$ &  $1$ & $-1$ & -    & $-1$ & $0.72$ &    $4.847^{\circ}$ &   $0.69202$ d \\
 $3$ & $-1$ & $-1$ & -    &  $1$ & $0.70$ &   $-0.335^{\circ}$ &  $-0.69170$ d \\
 $4$ & $-1$ &  $1$ & -    &  $1$ & $0.70$ &  $175.154^{\circ}$ &  $-0.69202$ d \\
 $5$ & -    & -    & $-1$ &  $1$ & $0.59$ &    $4.581^{\circ}$ & $-87.81766$ d \\
 $6$ &  $1$ &  $2$ & -    & $-1$ & $0.56$ &   $-1.918^{\circ}$ &   $0.69153$ d \\
 $7$ &  $1$ & $-2$ & -    & $-1$ & $0.56$ &    $7.100^{\circ}$ &   $0.69219$ d \\
 $8$ & $-1$ & $-2$ & -    &  $1$ & $0.55$ & $-178.082^{\circ}$ &  $-0.69153$ d \\
 $9$ & $-1$ &  $2$ & -    &  $1$ & $0.55$ &  $172.900^{\circ}$ &  $-0.69219$ d \\
$10$ & -    & -    &  $1$ & $-1$ & $0.43$ &  $175.419^{\circ}$ &  $87.17665$ d \\
$11$ &  $1$ & -    & -    & $-1$ & $0.34$ &    $2.593^{\circ}$ &   $0.69186$ d \\
$12$ & $-1$ & -    & -    &  $1$ & $0.33$ &  $177.407^{\circ}$ &  $-0.69186$ d \\
 \tableline
\end{tabular}
\end{table}

\end{document}